\documentclass[preprint,unsortedaddress,amsmath,amssymb,superscriptaddress]{revtex4-1}
\usepackage{amsmath}
\usepackage{graphicx}
\usepackage{bm}
\newcommand{\ef}{$\varepsilon$\textsubscript{F}\,}
\newcommand{\dpoc}{$\Delta_{\rm pocket}$\,}

\usepackage{hyperref}
\hypersetup{
    colorlinks=true, 
    linktoc=all,     
    linkcolor=blue,
    citecolor=blue,
    urlcolor=black,
    breaklinks=true
    }  
\usepackage[all]{hypcap}

\renewcommand{\vec}[1]{\boldsymbol{\mathbf{#1}}}

\usepackage{newfloat}

\DeclareFloatingEnvironment[name={Extended Data Figure}]{suppfigure}

\let\%\relax 
\DeclareFontFamily{OT1}{pxr}{}
\DeclareFontShape{OT1}{pxr}{m}{n}{<->pxr}{}
\DeclareSymbolFont{letA}{OT1}{pxr}{m}{n}
\DeclareMathSymbol{\%}{0}{letA}{`\%}
\usepackage{lineno}
\renewcommand{\baselinestretch}{1.3}

\begin{document}


\title{BCS-BEC crossover driven by small Fermi pockets of a high-$T_{\rm c}$ cuprate superconductor}
\author{Junhyeok~Jeong}
\affiliation{Institute for Solid State Physics, The University of Tokyo, Kashiwa, Chiba 277-8581, Japan}

\author{Yamato~Enomoto}
\affiliation{Department of Applied Electronics, Tokyo University of Science, Tokyo 125-8585, Japan}

\author{Yoshimitsu~Kohama}
\affiliation{Institute for Solid State Physics, The University of Tokyo, Kashiwa, Chiba 277-8581, Japan}

\author{Tomotaka~Nakayama}
\affiliation{Department of Applied Electronics, Tokyo University of Science, Tokyo 125-8585, Japan}

\author{Kotaro~Ando}
\affiliation{Department of Applied Electronics, Tokyo University of Science, Tokyo 125-8585, Japan}

\author{Kifu~Kurokawa}
\affiliation{Institute for Solid State Physics, The University of Tokyo, Kashiwa, Chiba 277-8581, Japan}

\author{Soonsang~Huh}
\affiliation{Institute for Solid State Physics, The University of Tokyo, Kashiwa, Chiba 277-8581, Japan}

\author{Zhuo~Yang} 
\affiliation{Institute for Solid State Physics, The University of Tokyo, Kashiwa, Chiba 277-8581, Japan}

\author{Toshihiro~Nomura} 
\affiliation{Department of Physics, Faculty of Science, Shizuoka University, Shizuoka 422-8529, Japan}

\author{Matthew~D.~Watson}
\affiliation{Diamond Light Source Ltd, Harwell Science and Innovation Campus, Didcot, OX110DE, U.K.}

\author{Timur~K.~Kim}
\affiliation{Diamond Light Source Ltd, Harwell Science and Innovation Campus, Didcot, OX110DE, U.K.}
   
\author{Cephise~Cacho}
 \affiliation{Diamond Light Source Ltd, Harwell Science and Innovation Campus, Didcot, OX110DE, U.K.}

\author{Chun~Lin}
\affiliation{Stanford Synchrotron Radiation Lightsource, SLAC National Accelerator Laboratory, 2575 Sand Hill Road, Menlo Park, CA 94025, U.S.A.}
 
  \author{Makoto~Hashimoto}
 \affiliation{Stanford Synchrotron Radiation Lightsource, SLAC National Accelerator Laboratory, 2575 Sand Hill Road, Menlo Park, CA 94025, U.S.A.}
 
 \author{Donghui~Lu}
 \affiliation{Stanford Synchrotron Radiation Lightsource, SLAC National Accelerator Laboratory, 2575 Sand Hill Road, Menlo Park, CA 94025, U.S.A.}
 
 \author{Shiro~Sakai} 
\affiliation{Faculty of Science and Technology, Sophia University, Tokyo 102-8554, Japan}

\author{Takami~Tohyama}
\affiliation{Department of Applied Physics, Tokyo University of Science, Tokyo 125-8585, Japan}

\author{Kazuyasu~Tokiwa} 
\email[Corresponding author: Mail address: ]{tokiwa@rs.tus.ac.jp}
\affiliation{Department of Applied Electronics, Tokyo University of Science, Tokyo 125-8585, Japan}

\author{Takeshi~Kondo}
\email[Corresponding author: Mail address: ]{kondo1215@issp.u-tokyo.ac.jp}
\affiliation{Institute for Solid State Physics, The University of Tokyo, Kashiwa, Chiba 277-8581, Japan}
\affiliation{Trans-scale Quantum Science Institute, The University of Tokyo, Bunkyo-ku, Tokyo 113-0033, Japan}

\date{\today}

\maketitle

\noindent\textbf{Abstract}

\textbf{
Fermi arcs observed in underdoped cuprates have sparked debate over whether they represent segments of a large Fermi surface or small Fermi pockets. This ambiguity has long hindered their classification as either the conventional Bardeen-Cooper-Schrieffer (BCS) regime or the strongly coupled Bose-Einstein condensation (BEC) crossover limit. Here, using angle-resolved photoemission spectroscopy and quantum oscillations, we demonstrate the coexistence of a small Fermi pocket and a large superconducting gap in the clean inner CuO$_2$ layers of the four-layer cuprate Ba$_2$Ca$_3$Cu$_4$O$_8$(F,O)$_2$. This coexistence constitutes a hallmark of the BCS-BEC crossover and has remained elusive for decades.
Despite the presence of antiferromagnetic (AF) order, the superconducting gap in the small pocket is remarkably large, yielding a gap-to-Fermi energy ratio ($\Delta_{\text{pocket}}/\varepsilon_{\rm F} \sim 0.6$) and a critical-to-Fermi temperature ratio ($T_{\rm c}/T_{\rm F} \sim 1.3$) that reach the theoretical upper bound for two-dimensional superconductivity. Unexpectedly, this BCS-BEC crossover emerges not as the carrier density decreases but as it increases, abruptly within a narrow doping range of less than 1$\%$. These results provide a long-sought microscopic foundation for the $d$-wave pairing mechanism in doped AF-Mott insulators.}
\newpage
\noindent\textbf{Introduction}

The distinction between a large Fermi surface and a small Fermi pocket with carrier density $1 + p$ and $p$, respectively, remains one of the most fundamental unsolved issues in high-$T_{\rm c}$ cuprates \cite{badoux2016,putzke2021}. The carrier density is traditionally estimated from the inverse Hall coefficient via the Drude formula ($R_{\rm H} = 1/ne$, where $n$ and $e$ are the carrier density and elementary charge), which assumes an isotropic, closed Fermi surface within the Boltzmann transport framework. This “effective” carrier density offers a simplified view of the electronic phase diagram. However, this picture stands in apparent contradiction to angle-resolved photoemission spectroscopy (ARPES) results, which reveal not a closed Fermi surface but rather an anomalous Fermi arc \cite{Damascelli}. Because the Fermi arc may represent a fragment of either a large surface or a small pocket, the carrier density in the arc state cannot be determined as either $1 + p$ or $p$ \cite{doiron2007,kyleshen2005}. Notably, a large Hall coefficient (indicative of small pockets) has been explained by a theory invoking anisotropic scattering on a large Fermi surface~\cite{Kontani}. This theory may also be relevant to recent angle-dependent magnetoresistance (ADMR) results~\cite{Ramshaw} presented as evidence for hole pockets. The longstanding issue, expressed as “a large Fermi surface vs. a small Fermi pocket”, is apparently not easy to resolve. 

Similarly, the presence of the Fermi arc obscures the Fermi energy ($\varepsilon_{\rm F}$) value, which is estimated from the corresponding band bottom or top. When the Fermi arc is assumed as a fragment of the large Fermi surface centered at ($\pi$, $\pi$), $\varepsilon_{\rm F}$ is considered to be a substantially large value of approximately 1 eV \cite{kramer2019}. In such a case, the pairing strength $\Delta / \varepsilon_{\rm F} \sim d_{\rm p}/\xi$ (where $\Delta$ is the energy gap), which describes the ratio between the distance between pairs $d_{\rm p}$ and the pair size $\xi$ (or coherence length), is estimated to be very small (around 0.03 for a typical gap size of $\sim 30$~meV), placing it within the weak-coupling regime. In contrast, if the Fermi arc is part of a small pocket, $\varepsilon_{\rm F}$ should be much smaller (at least an order of magnitude lower), resulting in a relatively large $\Delta / \varepsilon_{\rm F}$, which could lie in the BCS-BEC crossover regime. This stark discrepancy, arising from the interpretation of the Fermi arc, presents a major challenge in modeling the pairing mechanism in cuprates.

Most prior studies have focused on single- and double-layer cuprates due to their structural simplicity and relative ease of crystal growth. However, in these systems, disorder is inevitably introduced into the CuO$_2$ superconducting layers owing to their proximity to the dopant layers, which contain random potentials associated with elemental vacancies~\cite{McElroy2005}; this produces inhomogeneous electronic states, as revealed by scanning tunneling microscopy~\cite{McElroy2005,Pan_STM}. Such disorder leads to a serious mismatch between idealized theoretical models and real materials, as most theories neglect the effects of disorder. Without resolving this discrepancy, a breakthrough in formulating a microscopic theory of pairing may not be achieved. In particular, elucidating the electronic nature of the heavily underdoped regime is crucial, as this is the region where superconducting pairing begins upon doping an antiferromagnetic Mott insulator. Nonetheless, this region is highly sensitive to disorder owing to the weak screening effect stemming from the low carrier density.

A nearly disorder-free CuO$_2$ environment, which offers a solution, is realized in the inner layers of multilayer cuprates. In the multilayer configuration, these inner layers are spatially separated from the dopant layers, which are the primary source of random Coulomb potentials and lattice distortions. The outer planes effectively screen these disorder potentials, leading to a substantial suppression of both structural distortion and electronic inhomogeneity in the inner CuO$_2$ planes. In particular, the systematic nuclear magnetic resonance NMR studies of multilayer cuprates have demonstrated that in multilayer systems with three or more CuO$_2$ layers per unit cell ($n \geq 3$, $n$ is the number of CuO$_2$ layers per unit cell), the inner CuO$_2$ planes (IPs) exhibit significantly narrower NMR linewidths than not only the outer planes (OPs) but also single- and double-layer cuprates ($n = 1, 2$) \cite{Mukuda2012,shimizu2009}. This sharp spectrum directly indicates a substantially reduced degree of both lattice and electronic disorder in the IPs. Notably, the highest superconducting critical temperature ($T_{\rm c}$) among all known materials at ambient pressure has been observed in triple-layer cuprates, where such clean inner CuO$_2$ layers are present \cite{Schilling_Hg1223,Mukuda2012}. Investigating the inner layers of multilayer cuprates is thus likely the key to unlocking a major breakthrough in understanding high-$T_{\rm c}$ superconductivity; nevertheless, such studies have long been hindered by the difficulty of growing high-quality single crystals.

The argument of “a large Fermi surface vs. a small Fermi pocket” appeared to be resolved by the recent observation of small Fermi pockets in the high-quality single crystals of multilayer cuprates containing five or six CuO$_2$ layers per unit cell \cite{Kunisada2020,Kurokawa2023}. However, the superconducting gap for these pockets was very small ($\sim$~5~meV). Notably, antiferromagnetic (AF) order coexists with superconductivity in multilayer cuprates with $n \geq 3$ \cite{Shimizu2012,Mukuda2012}. While AF order may assist in stabilizing Fermi pocket formation, it may also compete with superconductivity, thereby reducing the superconducting gap \cite{Demler1993}. Thus, the small superconducting gap observed in the pocket band can be expected. Yet, there remains a possibility that the gap magnitude was small because the doping level is close to the edge of $T_{\rm c}$-dome \cite{Kurokawa2023}. Nonetheless, tuning doping levels in single crystals of multi-layered systems is highly challenging, and in fact, there has been no such report for the five- and six-layer compounds.

In this article, we focus on the four-layer Ba$_2$Ca$_3$Cu$_4$O$_8$(F,O)$_2$ (F0234), consisting of only one set of double inner planes per unit cell. This compound allows for a higher doping level than its five- or six-layer counterparts, owing to fewer inner layers being doped by dopant layers. Here we demonstrate that small Fermi pockets can host an substantially large superconducting gap, $\Delta_{\rm pocket} \sim$~30~meV at the pocket tip and $\Delta_{0} \sim$~60~meV extrapolated to the antinode. This gap ranks among the largest reported in cuprates and is comparable to that of HgBa$_2$Ca$_2$Cu$_3$O$_{8+\delta}$~\cite{Horio2025} with the highest $T_{\rm c}$ in cuprate ($T_{\rm c}$ = 130 K). The strong pairing strength $\Delta_{\rm pocket} / \varepsilon_{\rm F} \sim 0.6$, along with simultaneously observed pairing pseudogap and Bogoliubov flat-band, all place the inner-layer electronic state in the strong-coupling BCS-BEC crossover regime. Strikingly, the BCS-to-BEC crossover emerges not upon reducing, but rather upon increasing the carrier density, contrary to conventional expectations, and it occurs abruptly within a very narrow doping window. These findings offer the experimental insight into the mechanism of electron pairing that emerges upon doping an antiferromagnetic Mott insulator in a nearly disorder-free CuO$_2$ plane of cuprates, thereby shedding light on the fundamental nature of high-$T_{\rm c}$ superconductivity.\\

\noindent\textbf{Results}

A small Fermi pocket has not been observed in four-layer cuprates to date. Here, we present its observation using F0234 with different doping levels of $T_{\rm c}$ = 71K, 74K, and 78K (UD71K, UD74K, and UD78K, respectively; see Fig. 1b). Figures 1c-e show the Fermi surface mapping for the three samples measured by ARPES. The spectral intensities are taken slightly below the Fermi level to avoid spectral suppression due to the superconducting gap. We found a small Fermi pocket, along with a typical Fermi arc characteristic of the underdoped cuprates, in all three samples. In multilayered cuprates, each layer forms a Fermi surface independently of other layers. Therefore, the small pocket and the Fermi arc are each assigned to IPs and OPs, as depicted in Fig. 1a. The Fermi pocket is also confirmed by the de Haas-van Alphen oscillations (Figs. 1g-i; see Supplementary NOTE 2 for more details). Importantly, the oscillation frequencies are nearly identical around 380 T for all the samples, indicating that the doping levels $p$ (the proportion of the pocket size to the whole Brillouin zone) in the IP are nearly the same ($p\sim5.5\%$) within measurement error of $\pm\sim0.3\%$. This is also confirmed in Fig. 1f, which plots the Fermi momentum ($k_{\text{F}}$) points extracted from the ARPES spectra. Both the pockets and arcs for the three samples overlap with each other without a sizable difference. In particular, we find that the Fermi pocket of $p$=$5.5\%$ (a black oval) matches well with the ARPES data. 

The slight doping differences among the three samples are elucidated by variations in their effective masses. Previous reports have shown that the effective mass increases with doping in the underdoped regime, regardless of their different underlying mechanisms between charge order and Mott physics \cite{Ramshaw2015,Kunisada2020,Kurokawa2023}. We therefore posit a similar doping-dependent behavior in the four-layer compounds as well. The effective mass ($m^*$) is obtained from the temperature dependence of the quantum oscillation amplitude using the Lifshitz-Kosevich formula (insets of Figs. 1g-i). As depicted in Fig. 1j, we find an increasing trend of $m^*$ with increasing $T_{\rm c}$. We note that the variation of the effective mass observed in quantum oscillation measurements is subtle, ranging from $m^* \sim 0.82~m_0$ in UD71K to $\sim 0.91~m_0$ in UD78K (Figs. 1g-i). In addition, since both the inner and outer CuO$_2$ layers of all three samples remain in the underdoped regime (Supplementary NOTE 5), a higher $T_{\rm c}$ implies a higher hole concentration. Taken together, these observations indicate the slight increase in carrier density $p$ from UD71K to UD78K. We emphasize, however, that the doping levels of the three samples vary within a very narrow range (approximately 0.6$\%$) around 5.5$\%$, as estimated from the measurement error ($\pm\sim0.3\%$) of the quantum oscillation frequency.

As demonstrated in Supplementary NOTE 14 and Fig. S13, the expected change in band dispersion slope associated with the mass variation among three samples is substantially smaller than the intrinsic width of the ARPES spectra, making it impractical to resolve this difference directly by ARPES. Nevertheless, the band dispersion derived from the effective masses obtained from quantum oscillations is in excellent agreement with that observed by ARPES, highlighting the consistency between the two techniques.

A small Fermi pocket corresponds to small $\varepsilon_{\rm F}$, which is essential for achieving a strong pairing strength ($\Delta_{\rm pocket} / \varepsilon_{\rm F}$). To accurately characterize the band structure of IP, we perform band-selective measurements, where contamination of the ARPES signal from the outer CuO$_2$ plane (OP) is suppressed by exploiting the matrix element effect in photoemission (see Figs. 4a,b for mappings). Figures 1k,l show the ARPES band map and extracted band dispersion of UD71K at the superconducting state. The pocket band becomes more prominent away from the nodal cut (or diagonal cut), so we select the vertical cut (along magenta arrow in Fig. 1n) from ($\pi/2$, 0) to ($\pi/2$, $\pi$), where the superconducting gap is small enough to minimize its effect on the band determination. By fitting the occupied band determined by ARPES, we estimate the band top energy (\ef) as $\sim 50$ meV (see Supplementary NOTE 3 for UD74K and UD78K). We also confirm the consistent value of \ef via the measurements of a band along the AF zone boundary (AFZB, purple arrow) for UD74K (Fig. 1m). The data of Fig. 1m were taken at a high temperature (230 K) above $T_{\rm c}$ and divided by the resolution-broadened Fermi-Dirac (FD) function to trace the band dispersion up to \ef.  We note that a sharp parabolic dispersion with well-defined spectral features, as observed along the AFZB here, is never seen for a band with a Fermi arc. Importantly, \ef $\sim 50$ meV persists across all three samples.

Next, we measure the superconducting gap along the small pocket. Figures 2a-f present the ARPES spectra for the three samples and their curvature plots \cite{zhang2011}, which enhance the visibility of peak positions. 
The spectra were measured along the momentum cut crossing the pocket tip (Fig. 2g, also see Supplementary NOTE 12), where the largest energy gap (\dpoc) opens in the small pocket. We found that \dpoc for UD71K, UD74K, and UD78K are around 15 meV, 19 meV, and 30 meV, respectively. 
Considering their minimal carrier differences, it is remarkable that the gap size nearly doubles from UD71K to UD78K. 
Furthermore, the magnitude of this gap is significantly large, albeit with the low carrier concentration of $p\sim5.5\%$.
In Fig. 2j, we compare the energy distribution curves (EDCs) between the IP of the current sample and the optimally doped double-layered Bi$_2$Sr$_2$CaCu$_2$O$_{8+\delta}$(Bi2212) with a higher $T_{\rm c}$ of 92 K \cite{Kondo_Bi2212}, which is the most well-studied cuprate by ARPES. 
We find that the gap magnitude is about twice as large in IP. 
In Fig. 2l, we plot the gap size as a function of $d$-wave form $|\cos(k_x)-\cos(k_y)|/2$ (Supplementary NOTE 13). The SC gap extrapolated to the antinode ($\Delta_0$) reaches $\sim 60$ meV, which is one of the largest reported in ARPES studies of cuprates \cite{Kondo,Zhong2018,Vishik2012,Zhong2022Lifshitz,Nakayama2007,Ideta2012,Kunisada2017,Vishik2014} and is even comparable to that of HgBa$_2$Ca$_2$Cu$_3$O$_{8+\delta}$~\cite{Horio2025} with the highest $T_{\rm c}$ in cuprate ($T_{\rm c}$ = 130 K).

We further find that the gap magnitude of IP is about twice that of OP, as demonstrated in Fig. 2i by overlaying EDCs for IP and OP. 
Although the OP, being adjacent to the dopant layers, has a higher carrier density than the IP, it remains within the underdoped regime (Supplementary NOTE 5 and 6). Via the investigation of single- and double-layer cuprates \cite{Anzai2013,Zhong2018,Vishik2012}, the superconducting gap in the underdoped regime is almost constant or decreases with lower carrier density. Hence, the significant difference in gap magnitude between IP and OP cannot be understood by their doping level difference. We also point out that NMR measurements identified the presence of AF order in the IP for the current compound with similar doping levels to our samples, while it is negligible in OP \cite{shimizu2009}. The AF order, therefore, does not compete with superconductivity, but rather intimately coexists. Interestingly, the emergence of a large superconducting gap in a small pocket band indicates that a high-$T_{\rm c}$ superconductivity does not rely on the contributions from the antinodal electronic states, where the gap magnitude and superfluid density partition are maximal. 

A large superconducting gap ($\Delta$) in a small pocket with a small Fermi energy naturally highlights the importance of $\Delta/\varepsilon_{\rm F}$~\cite{Chen2005,chen_crossover}, a dimensionless parameter that defines the pairing strength in superconductors. Figure~2k plots the pairing strength $\Delta_{\rm pocket}/\varepsilon_{\rm F}$, calculated using the experimentally observed superconducting gap at the pocket tip ($\Delta_{\rm pocket}$) and $\varepsilon_{\rm F}$ for the three samples. We obtain remarkably large values ranging from 0.3 to 0.6, which are several orders of magnitude greater than those of conventional BCS-type superconductors, and even exceed those of any cuprate compounds~\cite{Uemura2004}. The pairing strength in the IP of our four-layer cuprates is notably comparable to, or even exceeds, that of bulk and monolayer FeSe~\cite{Liu2012,Lubashevsky2012,Kasahara2014,rinott2017}, Li$_x$ZrNCl~\cite{Nakagawa2021}, and magic-angle bilayer graphene~\cite{matbg_cao,matbg_oh}, which are all reported to reside in the BCS-BEC crossover regime (Supplementary NOTE 9).

The core concept of BCS-BEC crossover lies in the formation of tightly bound molecule-like bosonic pairs, which exhibit distinct characteristics compared to classical BCS superconductors. One of the main features is a pseudogap that opens above $T_{\rm c}$, indicating the pair formation at a higher temperature than that for pair condensation. Here, we note that the pairing pseudogap arises as a precursor to superconductivity \cite{Chen2005,chen_crossover,randeria2014}, and it should be distinguished from the pseudogap competing with superconductivity that develops near ($\pi,0$) in a wide doping range of cuprates \cite{tanaka2006,Kondo,Hashimoto}. 

Figures 3a-c show the temperature evolution of EDCs at $k_{\text{F}}$ for the pocket tip ($k_{\text{F}}^{\text{tip}}$, see Fig. 2g).
The EDCs are symmetrized about the Fermi level to eliminate the Fermi cut-off, allowing for the visualization of gap opening or closing. 
The spectra at low temperatures well below $T_{\rm c}$ exhibit coherent peaks and the superconducting gap.
Intriguingly, a spectral gap persists even above $T_{\rm c}$ in all samples. For UD71K and UD74K (Figs. 3a,b, respectively), well-defined peaks are observed above $T_{\rm c}$. This allows us to determine the closing temperature ($T_{\rm pair}$) of the pairing pseudogap:  $T_{\rm pair}$$\sim80$ K and $T_{\rm pair}$$\sim90$ K for UD71K and UD74K, respectively. In contrast, the superconducting peak for UD78K (Fig. 3c) becomes ill-defined with increasing temperature and the gapped state with spectral depletion around the Fermi level continues up to temperatures above $T \sim165$ K. This indicates that the pseudogap which competes with superconductivity \cite{tanaka2006,Kondo,Hashimoto} develops in UD78K due to its slightly higher doping level compared to the other two samples. 

To precisely determine $T_{\text{pair}}$ for UD78K, we analyze the spectral weight filling within the gap for symmetrized EDCs (Fig. 3d). In Fig. 3e, we plot the spectral weight around the Fermi level (gray region in Fig. 3d) as a function of the temperature. We find that, around 100 K, the rate of increase of the spectral weight changes to a $T$-linear behavior, which is characteristic of the competing pseudogap \cite{KondoNP}. From the onset temperature of the deviation, we determine $T_{\rm pair}$ to be $\sim 100$ K, which is significantly higher than $T_{\rm c}$. Figure 3f plots the ratio of $T_{\rm pair}$ to $T_{\rm c}$ for three samples, showing that the temperature for the preformed pair is raised with an increase of the doping level. 

Another notable feature of the BCS-BEC crossover is a Bogoliubov flat band, which signifies the presence of spatially localized pair states. The pair size $\xi$ (or coherence length) is roughly given by $\xi$ $\sim$ $1/\delta k_{\rm F}$ $\sim$ $\hbar v_{\rm F}/2 \Delta$ with the momentum span of the flat portions of Bogoliubov bands ($\delta k_{\rm F}$) expressed as the ratio between the SC gap and the Fermi group velocity ($v_{\rm F}$). On the other hand, the average inter-pair distance $d_{\rm p}$ is of the order of 1/$k_{\rm F}$ $\sim$ $\hbar v_{\rm F}/2 \varepsilon_{\rm F}$ with the Fermi wave number ($k_{\rm F}$). Therefore, the pairing strength is expressed as their ratio: $d_{\rm p}$/$\xi$ $\sim$ $\delta k_{\rm F}/k_{\rm F}$ $\sim$ $\Delta/\varepsilon_{\rm F}$. The flat band that spans from +$k_{\rm F}$ to -$k_{\rm F}$ in the pocket indicates a large $\Delta_{\rm pocket} / \varepsilon_{\rm F}$ of around 1 (that is, the pair size and the distance of pairs becomes comparable), corresponding to the BCS-BEC crossover regime. 

The Bogoliubov flat band has been observed in FeSeTe \cite{rinott2017} and FeSeS \cite{hashimoto2020} systems, providing evidence of the BCS-BEC crossover. Here, we demonstrate that four-layer cuprates also have such a flat band. We employ the matrix element effect in ARPES to 
suppress the spectral signals of the OP band, enabling a detailed examination exclusively of the IP band structure. Figure 4a shows the Fermi surface mapping with light polarization aligned along the sample $a$-axis. We found an adequate momentum region where only the pocket band is observed in the 2nd Brillouin zone. In Figs. 4c-e, we extract the ARPES band dispersions along the momentum Cut 1 (white line in Fig. 4a), which crosses the pocket center horizontally for the three samples. Overall band dispersions are similar in all samples, reflecting their small differences in doping levels.

In Fig. 4h, we present EDCs extracted from the ARPES map for UD78K (Fig. 4e) with an offset for clarity. We find a Bogoliubov flat band spanning from +$k_{\rm F}$ to -$k_{\rm F}$, which is consistent with a strong pairing strength $\Delta_{\rm pocket} / \varepsilon_{\rm F}$. A clear difference among three samples is seen around the pocket center in the prominence of the spectral peak at the pocket center ($\pi/2, \pi/2$), as shown in Fig. 4i. A coherent peak appears in UD78K (red curve in Fig. 4i), whereas it is less evident in the other two samples (UD71K and UD74K). Overall, the Bogoliubov flat band is most pronounced in UD78K.
Notably, the Bogoliubov quasiparticle peak is not expected at ($\pi/2, \pi/2$), which lies along the nodal direction. A likely cause for a sharp peak is strong coupling with a collective mode, as evidenced by a peak-dip-hump structure in our data (red curve in Figs. 4i,k). The mode energy, estimated to be $\sim$30-40 meV from the peak-to-dip energy spacing, corresponds to a typically observed mode in cuprates~\cite{Feng_kink,Zhou_LSCO,Kunisada2017,Zhou_kink}. Our results thus indicate that mode coupling, which is enhanced in the superconducting state, facilitates the generation of a continuous Bogoliubov flat band without a disconnection across the nodal ($\pi/2, \pi/2$) point.  Furthermore, the flat band observed is apparently flatter than a conventional BCS-type Bogoliubov band. This implies that mode coupling renormalizes the band and reinforces the flatness (Supplementary NOTE 11), thereby enhancing the spatial localization of paired electrons.

To confirm that this conclusion is not sensitive to the matrix element effect, we measured the ARPES data for samples rotated by 45 degrees in the azimuth angle, maintaining light polarization with respect to the analyzer. The obtained Fermi surface map is displayed in Fig. 4b. We find another adequate momentum region in a different 3rd Brillouin zone, where only the pocket signals are pronounced. For two samples, UD74K and UD78K, the band dispersions are extracted along the diagonal direction of Cut 2, which crosses the pocket center diagonally (or along AFZB), as represented in Fig. 4b with a white line. Again, we find a coherent peak at the pocket center in UD78K, which is instead unclear in UD74K (Fig. 4k). A Bogoliubov flat band is also reproduced in Fig. 4j, which plots EDCs for the ARPES map of Fig. 4g. These consistent results, which are robust against the matrix element effect, clearly demonstrate that the flat band is an intrinsic property and that UD78K is particularly closer to the BEC regime. 

Our findings are summarized in Fig. 5. We found that a small pocket and a large superconducting gap simultaneously emerge in cuprates.  
The AF order coexists with superconductivity in the inner plane that hosts a small pocket, yet its superconducting gap is significantly larger than that of the paramagnetic outer plane with a Fermi arc by a factor of two (Fig. 4l and Figs. 5a-c). This indicates that the AF order does not compete with the high-$T_{\rm c}$ conductivity but rather has an intimate relationship. 

In Figs. 5d-e, we illustrate an electronic phase diagram of multilayer cuprates based on our results. Although the doping variation we addressed is merely within 0.6~$\%$, a drastic variation occurs. Whereas the two temperatures of $T_{\rm pair}$ and $T_{\rm c}$ are going to merge towards the edge of the superconducting dome, the $T_{\rm pair}$ is much more rapidly increased at higher doping levels than the $T_{\rm c}$. This causes a clear discrepancy between the electron pairing and the superconducting condensation, highlighting a rapid approach toward the BEC regime in the inner layer of the four-layered cuprates. The well-acknowledged superconducting phase diagram in cuprates suggests that the BCS to BEC crossover occurs with decreasing carrier concentration from the overdoping to underdoping regimes~\cite{uemura1997,Uemura2004,Emery1995}. Interestingly, our data suggest a quite different trend: the crossover from BCS to BEC occurs with increasing doping level, contrary to the general expectation. 

Finally, we categorize the superconductivity of the four-layer cuprate in the Uemura plot \cite{Uemura2004}, which plots $T_{\rm c}$/\emph{T}\textsubscript{F} ratio of superconductors, where \emph{T}\textsubscript{F} is the Fermi temperature. For such categorization, it is crucial to confirm which layer dominantly determines the $T_{\rm c}$ value in our four-layer system. 
We obtained the following spectroscopic signatures indicating that the pocket band in IPs determines the $T_{\rm c}$ in the system. The extrapolated gap $\Delta_0$ in the IPs is approximately twice that in the OPs (Fig. 2l and Supplementary NOTE 6). Consistently, quasiparticle coherence peaks are markedly sharper and more pronounced for the IPs, implying reduced scattering and a more robust pair condensation (Supplementary NOTE 7). 
Most importantly, the OP gap closes at a temperature below the $T_{\rm c}$, whereas the IP gap persists up to temperatures higher than $T_{\rm c}$ (Supplementary NOTE 8) \cite{Sun2025,Jeong2025}. In light of these observations, we posit that the onset of superconductivity in our samples is determined by the inner layers. According to the Uemura plot \cite{Uemura2004}, most unconventional superconductors, including typical cuprates, heavy fermions, and iron chalcogenides, exhibit $T_{\rm c}$/\emph{T}\textsubscript{F}$\sim 0.05$, which is considerably distant from conventional BCS-type superconductors (Fig. 5f). Remarkably, the clean inner planes of the current four-layer cuprates show a much larger ratio ($T_{\rm c}$/\emph{T}\textsubscript{F}$\sim0.13$ for UD78K) than other unconventional superconductors, nearly reaching the theoretical upper bound expected in two-dimensional (2D) superconductivity ($T_{\rm c}$/\emph{T}\textsubscript{F}=0.125) \cite{Hazra2019}.\\


\noindent\textbf{Discussion and Outlook}

The coexistence of antiferromagnetic (AF) order and superconductivity is an intrinsic property of disorder-free underdoped cuprates, as consistently observed in the inner layers of multilayer systems with three or more CuO$_2$ planes per unit cell \cite{Mukuda2012,Shimizu2012,Kunisada2020,Kurokawa2023}. In such nearly ideal CuO$_2$ planes, superconductivity develops within a doped AF-Mott state while coexisting with static AF order, a regime that cannot be realized in single- or double-layer cuprates where disorder is unavoidable.

The crossover from BCS- to BEC-like behavior with increasing carrier density may appear counterintuitive. A natural explanation is that static AF order suppresses low-energy magnetic fluctuations, thereby stabilizing a more BCS-like regime even in a small Fermi pocket with low carrier density. As doping weakens the AF order, the magnetic exchange energy becomes comparable to the electronic kinetic energy associated with the pocket. Because the pocket has an intrinsically low $\varepsilon_{\rm F}$, even a slight increase in carrier density (only $\sim 0.6\%$) can substantially modify the balance between static order and dynamical fluctuations, enhancing the pairing scale relative to $\varepsilon_{\rm F}$ and pushing the system toward the BEC side of the crossover.

The inner CuO$_2$ planes thus lie in a regime where magnetic and kinetic energy scales are finely balanced. Such sensitivity suggests proximity to an antiferromagnetic quantum critical regime~\cite{Hashimoto_Science}, in which minute doping-induced changes can qualitatively alter the superconducting state. Our results demonstrate that in nearly ideal CuO$_2$ planes, tiny carrier tuning can switch superconductivity between BCS- and BEC-like regimes while preserving a well-defined small Fermi pocket. While other interactions, including electron-phonon coupling, may also contribute, magnetic fluctuations are expected to be strongly enhanced near AF instability and can influence the pairing scale.

The band structure in the antinodal region is characterized by low group velocity and therefore contributes significantly to the density of states and to the superfluid density in the superconducting state of cuprates. However, the pocket band lacks the contribution of antinodal states to pairing.  Our results, therefore, demonstrate that this seemingly advantageous feature of the electronic structure in high-$T_{\rm c}$ superconductivity is not essential for realizing a large superconducting gap and strong pairing strength. One possible scenario leading to this situation is that the pseudogap state, which competes with superconductivity \cite{tanaka2006,Kondo,Hashimoto}, is largely suppressed when a pocket band is formed, as its electronic states disperse down to $\sim$1~eV far from the Fermi level and the low-energy electronic states required for the development of the pseudogap are absent in the antinodal region \cite{Kunisada2020}. This could make electron pairing more stable in the pocket band state than in the Fermi arc state observed in underdoped single- and double-layer cuprates.

To clarify the unique position of cuprates in the broader context of BCS-BEC crossover research, we briefly compare them with other material systems. Fe-based superconductors have played a pioneering role as the only platform in which the BCS-BEC crossover has been directly identified via direct band observations~\cite{rinott2017, hashimoto2020}. However, its multi-orbital nature makes the BCS-BEC crossover scenario rather complicated, as the $\Delta$/\ef ratio decreases toward the BEC regime~\cite{hashimoto2020}. Layered nitrides \cite{Nakagawa2021}, which are doped semiconductors, serve as conceptually simple platforms with $\textit{s}$-wave pairing and minimal orbital complexity; in contrast, the exotic nature due to strong correlation is not induced. Organic superconductors \cite{suzuki2022} highlight the potential of low-dimensional correlated systems, though their surface sensitivity makes ARPES experiments challenging. Magic-angle twisted bilayer graphene \cite{matbg_cao,matbg_oh} provides a fascinating Mott-related system, albeit operating at energy and temperature scales more than an order of magnitude lower than those of cuprates. Within this diverse landscape, cuprates stand out as a uniquely powerful platform, combining a single-band, $\textit{d}$-wave symmetry, strong Mott-derived correlations, and high-temperature, large-energy scales. Moreover, the emergence of a well-defined pocket band in a clean inner CuO$_2$ layer enables direct comparison with theoretical models, offering an ideal setting for exploring the microscopic nature of the BCS-BEC crossover in strongly correlated systems. It is worth noting that the clean CuO$_2$ layer where the superconductivity coexists with AF order is likewise present in three-layer cuprates as well \cite{Mukuda2012,Shimizu2012,Oliviero}, which exhibits the highest $T_{\rm c}$ among all cuprate families with varying numbers of CuO$_2$ layers per unit cell \cite{Iyo2007}. This correspondence indicates that our results are not limited to a particular case, but reflect the intrinsic physics of clean CuO$_2$ planes, a key for elucidating the pairing mechanism of cuprate superconductors. Collectively, these features establish the multilayered cuprates as a uniquely positioned benchmark system for future experimental and theoretical investigations into strongly coupled superconductivity.

 \newpage

\noindent\textbf{Methods}

{\bf Samples:} Single crystals of underdoped Ba$_2$Ca$_3$Cu$_4$O$_{8}$(F,O)$_2$ (crystal structure is provided in Fig. 1a) with $T_c$=71 K, 74 K, and 78 K were grown at between 1000 $^\circ$C and 1200 $^\circ$C under a pressure of 4.5 GPa without an intentional flux. The starting composition for the crystal synthesis is Ba$_2$Ca$_{2.1}$Cu$_{3.1}$O$_{6.3}$F$_{2}$, which is known to be almost the same in the single crystals \cite{Tokiwa}. We have conducted X-ray diffraction measurements along the $c$-axis for all the sample pieces and confirmed that they are single crystals, not the mixtures of crystal domains with different numbers of CuO$_2$ layers per unit cell. Each single crystal piece from the batch exhibited a slight variance in the $T_{\rm c}$. We characterized every single crystal piece by measuring magnetic susceptibility (Fig. 1b) and classified them into the three doping levels, according to their $T_{\rm c}$. Laue image of the single crystal (Supplementary NOTE 1) shows a four-fold rotational symmetry with no indication of structural modulations. \\ 

{\bf ARPES measurements:} 
Laser-based ARPES data were accumulated using a laboratory-based system consisting of a Scienta R4000 electron analyzer and a 6.994 eV laser (the 6th harmonic of Nd:YVO$_4$ quasi-continuous wave) at the Laser and Synchrotron Research Center at Institute for Solid State Physics (ISSP), the University of Tokyo. The data presented are measured at 10 K. The overall energy resolution in the ARPES experiment was set to 1.6 meV. Synchrotron-based ARPES measurements were performed at the high-resolution branch (HR-ARPES) of the beamline I05 in the Diamond Light Source \cite{Hoesch_RSI_2017}, equipped with an MBS A-1 analyzer, and ARPES end station at Stanford Synchrotron Radiation Lightsource beamline 5-2, equipped with a Scienta DA30 analyzer. The data presented are measured at the photon energy of 55 eV and at the temperature of 7 K. The overall energy resolution was set to $\sim$10 meV in our experiments. In both the laser- and synchrotron-ARPES measurements, a typical cleavage method was used to get a clean surface of the samples: a top post glued on the crystal is hit {\it in situ} to obtain a flat surface suitable for the ARPES measurements. The cleavage plane has been confirmed by scanning tunneling microscopy to be along the F/O dopant layers \cite{Tokiwa}. The superconducting gap is obtained by fitting the ARPES spectra to the phenomenological model (Supplementary Information).
\\

{\bf  Quantum oscillation measurements:}
Torque magnetometry experiments were performed using a commercial piezoresistive cantilever (SEIKO PRC-120) \cite{Torque} in pulsed magnetic fields up to 60 T (36 ms pulse duration) at the International Megagauss Laboratory at ISSP, the University of Tokyo. The cantilever directly detects the magnetic torque ($\tau$) as the result of the anisotropic magnetization of the sample, $\tau=${\boldmath $M$}$\times${\boldmath $H$}, and the magnetic quantum oscillation known as the de Haas-van Alphen (dHvA) oscillation was observed. According to the Onsager relation, the quantum oscillation frequency $F$ is directly proportional to the extremal cross-sectional area of the Fermi surface perpendicular to the applied magnetic field, $S$, via $F=\hbar S/2\pi e$. In our measurements, the magnetic field was applied nearly parallel to the crystallographic $c$-axis, so that the oscillation frequency reflects the in-plane Fermi surface area (in the $ab$-plane).

In torque magnetometry measurement, the sign of the torque reverses when the magnetic field angle crosses the crystallographic $c$-axis. Therefore, we first determined the torque reversal angle through several low-field measurements. Based on this procedure, we aligned the sample $c$-axis nearly parallel to the external field direction, making the cross-sectional Fermi pocket area close to the area projected onto the cleave plane ($ab$-plane), which was used in the ARPES measurement. Here, the uncertainty of the field angle relative to the $c$-axis ($\theta$) is estimated to be approximately $\pm 1^{\circ}$. This level of angular uncertainty does not introduce a significant error in the estimation of the Fermi surface cross-sectional area with respect to the external field direction, since the angle-dependent quantum oscillation measurement revealed that the quantum oscillation frequency (cross-sectional area) of the small Fermi pocket follows a ($\cos{\theta}$) form, namely, two-dimensional-like behavior \cite{Kunisada2020}.
Supplementary Fig. 2 shows the magnetic torque signals after subtracting background, which was obtained by fitting a polynomial function to each curve of the raw data in the range of magnetic field between 37 and 60 T.\\



\newpage

\noindent\textbf{Acknowledgements}\\
We thank K. Mehlawat and T. Shitaokoshi for their help with the quantum-oscillation measurements, T. Yajima and J. Yamaura for technical assistance with the X-ray measurements, and T. Yamauchi for help with the magnetic-susceptibility measurements.
We thank Diamond Light Source for access to beamline I05 under proposals SI36822, SI30646, SI28930, SI25416, and Stanford Synchrotron Radiation Lightsource for access to beamline BL5-2 under proposal S-XV-ST-6368A that contributed to the results presented here.\\

\noindent\textbf{Fundings}\\
T.K. discloses support for the research of this work from JSPS KAKENHI [Grant numbers: JP21H04439, JP23K17351, JP25H01250, and JP25H01246], the Asahi Glass Foundation, MEXT Q-LEAP [Grant number: JPMXS0118068681], The Murata Science Foundation, The Mitsubishi Foundation, and Toray Science Foundation.\\

\noindent\textbf{Author contributions}\\
T.K. conceived and designed the project. 
J.J. performed the ARPES experiments with the help from K.K., S.H., M.D.W., T.K.K., C.C., C.L., M.H., D.L., and T.K. J.J. analyzed the data with the help from K.K. and T.K. Y.E., T.N., K.A., and K.T. grew the crystals, and J.J., Y.E., T.N., K.A., and K.T. conducted the sample characterization.
J.J., and Y.K. performed the quantum oscillations experiments with help from Z.Y. and T.N. J.J. analyzed the data with the help from K.K and Y.K.
J.J., S.S., T.T., K.T., and T.K. interpreted the data. 
All authors discussed the results, and J.J. and T.K. wrote the manuscript. 
T.K. and K.T. supervised the overall project.\\




\renewcommand{\baselinestretch}{1.15}
\begin{figure}[htbp]
\includegraphics [clip,width=16.5cm]{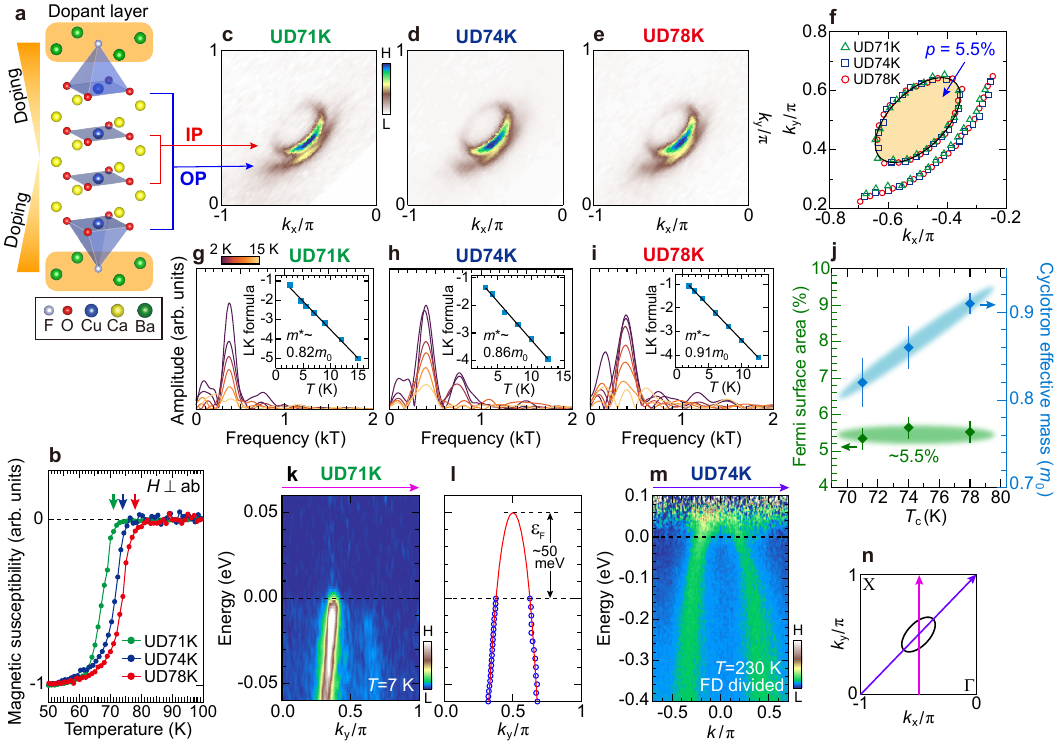}
\caption{\textbf{A small Fermi pocket observed by ARPES and quantum oscillation measurements in four-layer cuprates.} 
\textbf{a,} Crystal structure of Ba$_2$Ca$_3$Cu$_4$O$_{8}$(F,O)$_2$.
\textbf{b,} Magnetic susceptibility of single crystals (UD71K, UD74K, and UD78K) with $T_{\rm c}$ values determined from the onset of the Meissner effect (colored arrows). 
\textbf{c}-\textbf{e,} ARPES Fermi surface (FS) mappings at $T=7$ K for the three samples. The intensity map at -20 meV, integrated over $\pm 10$ meV, avoids the superconducting gap and reveals the FS shape. 
\textbf{f,} FSs determined by ARPES.
$k_{\text{F}}$ points were determined by identifying the momentum at which the superconducting gap reaches its minimum in EDCs. The black solid line corresponds to a small pocket with area $p=5.5\%$ estimated from quantum oscillation measurements. 
\textbf{g}-\textbf{i,} Fast Fourier transformation (FFT) of quantum oscillation spectra for UD71K, UD74K, and UD78K, respectively, measured from $T\sim2$ K to $\sim15$ K. 
Insets show the temperature dependence of FFT peak intensities with fits to the Lifshitz-Kosevich (LK) formula, yielding effective mass ($m^*$). 
 \textbf{j,} Small pocket area (green diamonds) and effective mass (blue diamonds) both obtained from quantum oscillation data are plotted against $T_{\rm c}$. 
 \textbf{k,} ARPES band dispersion of the pocket in UD71K, measured at $T=7$~K along the magenta arrow in (\textbf{n}). 
 \textbf{l,} Band dispersion determined from the peak positions of momentum distribution curves (MDCs) in (\textbf{k}). Red curve represents a tight-binding fit to the data (see Supplementary NOTE 4).
 \textbf{m,} ARPES band dispersion in UD74K along the purple arrow in (\textbf{n}), measured at $T=230$ K and divided by the resolution-convoluted Fermi-Dirac function.
 \textbf{n,} Momentum cuts for (\textbf{k,}\textbf{m}) (magenta and purple arrows, respectively) across the Fermi pocket (black line).} 
\end{figure}

\newpage

\begin{figure}[htbp]
\includegraphics [clip,width=16.5cm]{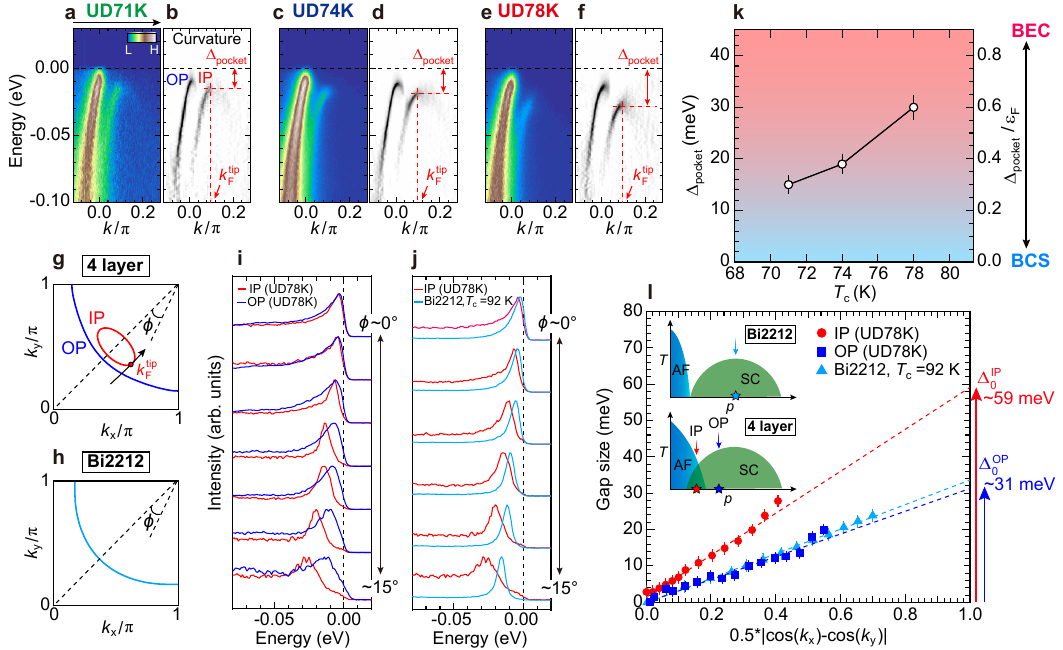}
\label{figure2}
\caption{\textbf{Superconducting gap and pairing strength of the pocket bands.} 
\textbf{a,} ARPES band dispersion of UD71K measured at $T=10$ K along the black arrow in (\textbf{g}), which cross the pocket tip ($k_{\text{F}}^{\text{tip}}$).
\textbf{b,} Curvature plot \cite{zhang2011} of (\textbf{a}): curvature of spectra to clarify peak positions. 
\textbf{c}-\textbf{f,} Similar data to (\textbf{a,}\textbf{b}), but for UD74K and UD78K.
 \dpoc is determined by EDC peaks at $k_{\text{F}}^{\text{tip}}$.
\textbf{g,} Schematic FSs for IP (red) and OP (blue) in the four-layer cuprate.
\textbf{h,} Schematic FS of optimally doped Bi2212. 
\textbf{i,} EDCs at $k_{\rm F}$ from $\phi=0^\circ$ (nodal point) to $15^\circ$ for IP (red) and OP (blue) in UD78K.
\textbf{j,} Similar data to (\textbf{i}), but for IP in UD78K (red) and optimally doped Bi2212 ($T_{\rm c}$=92 K) (light blue)~\cite{Kondo_Bi2212}. 
\textbf{k,} Superconducting(SC) gap at $k_{\text{F}}^{\text{tip}}$, \dpoc, and the corresponding pairing strength \dpoc/\ef plotted against $T_{\rm c}$. The error bars correspond to the estimated variances, considering measurement uncertainties of both the ARPES and quantum oscillation (see Supplementary Information for details).  
\textbf{l,} SC gap on each plane in UD78K and Bi2212 plotted against the $d$-wave function $|\cos(k_x)-\cos(k_y)|/2$. Gap values were determined from the symmetrized EDC peaks (see Supplementary Note 10 for a detailed analysis). 
$\Delta_0$ of IP and OP is extrapolated to the antinode. 
The inset shows schematic phase diagrams of Bi2212 and IP for the four-layer compound with doping levels indicated by stars and arrows. 
}
\end{figure}
\clearpage

\begin{figure}[htbp]
\includegraphics [clip,width=16.5cm]{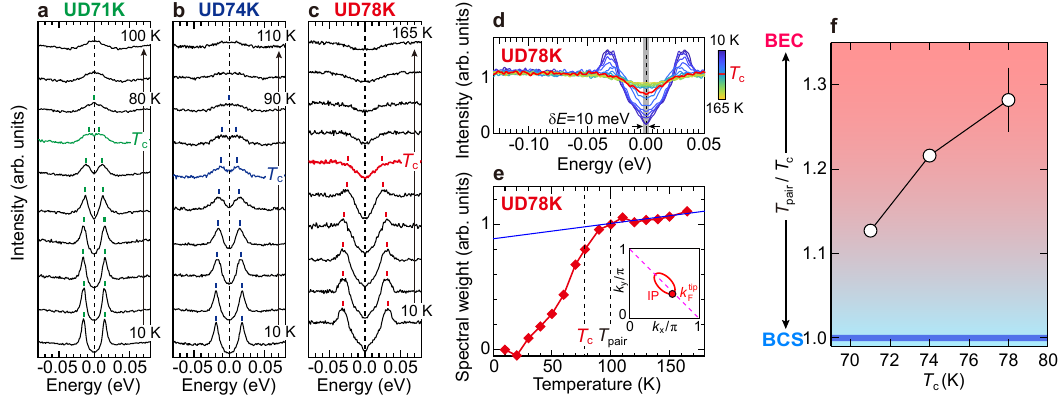}
\label{figure3}
\caption{\textbf{Pairing pseudogap in the pocket and its evolution with carrier doping} 
\textbf{a}-\textbf{c,} Temperature variation of symmetrized EDCs at the pocket tip for UD71K, UD74K, and UD78K, respectively. Spectra at $T_{\rm c}$ are colored. 
\textbf{d,} Same data as in (\textbf{c}), but plotted without offset.
\textbf{e,} Temperature dependence of spectral weight near the Fermi level, integrated over $\delta E = 10$ meV (gray region in (\textbf{d})). 
The onset temperature of pairing, $T_{\rm pair}$, is defined as the temperature where the high-$T$ linear behavior at high temperatures is deviated upon cooling.
\textbf{f,} $T_{\rm pair}$/$T_{\rm c}$ plotted against $T_{\rm c}$. 
The blue line at $T_{\rm pair}$/$T_{\rm c}$ = 1 indicates the BCS limit, where pair formation and condensation occur at the same temperature. The error bar for UD78K corresponds to the uncertainty in the spectral weight analysis (Supplementary NOTE 15).
}
\end{figure}
\clearpage

\begin{figure}[htbp]
\includegraphics [clip,width=16.5cm]{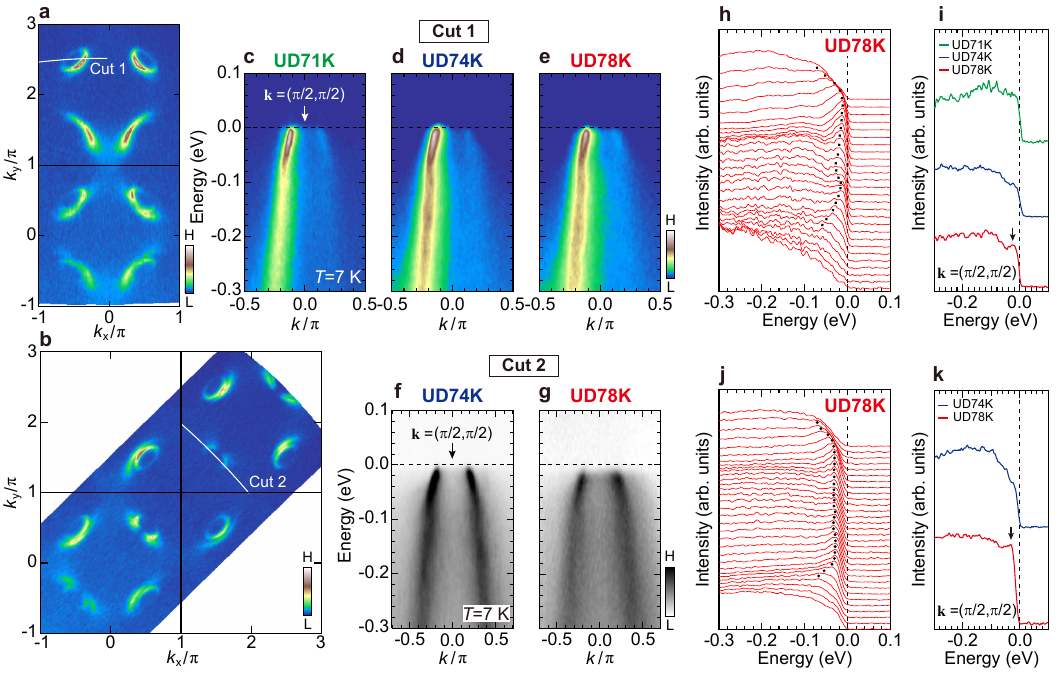}
\label{figure4}
\caption{\textbf{Bogoliubov flat band in the pocket.} 
\textbf{a,}\textbf{b,} ARPES FS maps of UD71K measured with different experimental geometries. 
\textbf{c}-\textbf{e,} ARPES band dispersion for UD71K, UD74K, and UD78K, measured along Cut 1 in (\textbf{a}). 
\textbf{f,}\textbf{g,} Similar data to (\textbf{c}-\textbf{e}), but measured along Cut 2 in (\textbf{b}). 
\textbf{h,} Full set of EDCs in (\textbf{e}). Peak positions (black circles) trace a Bogoliubov flat band.
\textbf{i,} EDCs at the pocket center ($\vec{k} = (\pi/2, \pi/2)$) extracted from (\textbf{c}-\textbf{e}). 
\textbf{j,} Same as (\textbf{h}), but for (\textbf{g}). These results consistently indicate a Bogoliubov flat band in the pocket.
\textbf{k,} Same as (\textbf{i}), but for (\textbf{f,}\textbf{g}). 
}
\end{figure}
\clearpage

\begin{figure}[htbp]
    \includegraphics [clip,width=16.5cm]{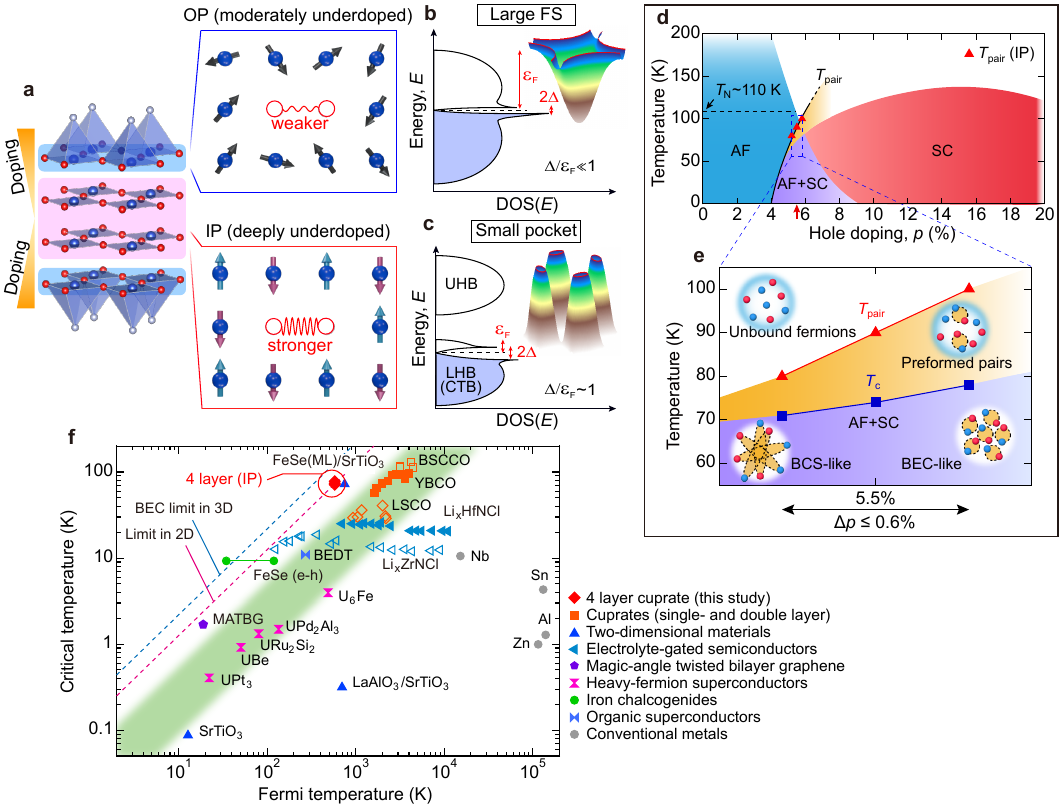}
    \label{figure5}
    \caption{}
\end{figure}
    \clearpage

\noindent\textbf{FIG. 5. Strongly bound pairs and phase diagram of the inner CuO$_2$ planes.} 
\textbf{a,} Schematic illustration of pairing strengths in each CuO$_2$ plane of four-layer cuprates. 
The IP forms a small Fermi pocket, is less doped and farther from optimal doping than the OP. The IP develops AF order, unlike the optimally-doped, paramagnetic OP with a large FS (or Fermi arc). Despite these features being seemingly unfavorable for stabilizing pairing, the SC gap of the IP is significantly larger than that of the OP. 
\textbf{b,} Schematic density of states (DOS) and band structure in the OP, featuring a large FS with a large Fermi energy \ef ($\sim$1~eV) and a small SC gap ($\sim$30~meV)~\cite{Damascelli,Ronning_waterfall}, resulting in a weaker pairing strength, $\Delta$/\ef $\sim$ 0.03 ($\ll1$). 
\textbf{c,} Schematic DOS and band structure in the IP: carriers are doped into the lower Hubbard band (LHB) or charge transfer band (CTB), generating a small Fermi pocket with a small Fermi energy \ef ($\sim$50~meV) and a large SC gap ($\Delta_{\rm pocket} \sim$~30~meV at the pocket tip and $\Delta_{0} \sim$~60~meV at the antinode). The SC gap and small \ef become comparable, yielding a stronger pairing strength $\Delta_{\rm pocket}/\varepsilon_{\rm F} \sim$~0.6, which is much larger than $\Delta$/\ef$\sim$ 0.03 in OP. 
\textbf{d,} Phase diagram of the inner CuO$_2$ planes. AF phase and superconducting dome are referenced from NMR studies on the same compound$~$\cite{Shimizu2012}. The purple region denotes coexisting AF order and superconductivity.
The red arrow marks the doping level of our samples ($p \sim 5.5\%$). Red triangles represent the $T_{\rm pair}$ values, evolving alongside the superconducting dome. The phase of preformed pairs is shaded in orange. 
\textbf{e,} Enlarged view of (\textbf{d}) (blue dashed box) focusing on the region around $p \sim 5.5\%$, summarizing our results and illustrating the BCS-BEC crossover from the BCS side (UD71K) to the BEC side (UD78K). 
\textbf{f,} Uemura plot: A logarithmic plot of critical temperature $T_{\rm c}$ versus Fermi temperature $T_{\text{F}}$ for various superconductors \cite{Kasahara2014,matbg_cao,matbg_oh,nakagawa2018,Nakagawa2021,Uemura2004,wang2012,lin2013}. The green shaded area indicates the region where most unconventional superconductors lie. The IPs of the four-layer cuprates, which form a small Fermi pocket, are marked by red diamonds; they reach the upper bound of the two-dimensional (2D) Berezinskii-Kosterlitz-Thouless (BKT) transition ($T_{\rm c}$/\emph{T}\textsubscript{F} = 0.125) \cite{Hazra2019}.

\end{document}



\title{Supplementary Information:\\BCS-BEC crossover driven by small Fermi pockets of a high-$T_{\rm c}$ cuprate superconductor}

\author{Junhyeok~Jeong}
\affiliation{Institute for Solid State Physics, The University of Tokyo, Kashiwa, Chiba 277-8581, Japan}

\author{Yamato~Enomoto}
\affiliation{Department of Applied Electronics, Tokyo University of Science, Tokyo 125-8585, Japan}

\author{Yoshimitsu~Kohama}
\affiliation{Institute for Solid State Physics, The University of Tokyo, Kashiwa, Chiba 277-8581, Japan}

\author{Tomotaka~Nakayama}
\affiliation{Department of Applied Electronics, Tokyo University of Science, Tokyo 125-8585, Japan}

\author{Kotaro~Ando}
\affiliation{Department of Applied Electronics, Tokyo University of Science, Tokyo 125-8585, Japan}

\author{Kifu~Kurokawa}
\affiliation{Institute for Solid State Physics, The University of Tokyo, Kashiwa, Chiba 277-8581, Japan}

\author{Soonsang~Huh}
\affiliation{Institute for Solid State Physics, The University of Tokyo, Kashiwa, Chiba 277-8581, Japan}

\author{Zhuo~Yang} 
\affiliation{Institute for Solid State Physics, The University of Tokyo, Kashiwa, Chiba 277-8581, Japan}

\author{Toshihiro~Nomura} 
\affiliation{Department of Physics, Faculty of Science, Shizuoka University, Shizuoka 422-8529, Japan}

\author{Matthew~D.~Watson}
\affiliation{Diamond Light Source Ltd, Harwell Science and Innovation Campus, Didcot, OX110DE, U.K.}

\author{Timur~K.~Kim}
\affiliation{Diamond Light Source Ltd, Harwell Science and Innovation Campus, Didcot, OX110DE, U.K.}
   
\author{Cephise~Cacho}
 \affiliation{Diamond Light Source Ltd, Harwell Science and Innovation Campus, Didcot, OX110DE, U.K.}

\author{Chun~Lin}
\affiliation{Stanford Synchrotron Radiation Lightsource, SLAC National Accelerator Laboratory, 2575 Sand Hill Road, Menlo Park, CA 94025, U.S.A.}
 
  \author{Makoto~Hashimoto}
 \affiliation{Stanford Synchrotron Radiation Lightsource, SLAC National Accelerator Laboratory, 2575 Sand Hill Road, Menlo Park, CA 94025, U.S.A.}
 
 \author{Donghui~Lu}
 \affiliation{Stanford Synchrotron Radiation Lightsource, SLAC National Accelerator Laboratory, 2575 Sand Hill Road, Menlo Park, CA 94025, U.S.A.}
 
 \author{Shiro~Sakai} 
\affiliation{Faculty of Science and Technology, Sophia University, Tokyo 102-8554, Japan}

\author{Takami~Tohyama}
\affiliation{Department of Applied Physics, Tokyo University of Science, Tokyo 125-8585, Japan}

\author{Kazuyasu~Tokiwa} 
\email[Corresponding author: Mail address: ]{tokiwa@rs.tus.ac.jp}
\affiliation{Department of Applied Electronics, Tokyo University of Science, Tokyo 125-8585, Japan}

\author{Takeshi~Kondo}
\email[Corresponding author: Mail address: ]{kondo1215@issp.u-tokyo.ac.jp}
\affiliation{Institute for Solid State Physics, The University of Tokyo, Kashiwa, Chiba 277-8581, Japan}
\affiliation{Trans-scale Quantum Science Institute, The University of Tokyo, Bunkyo-ku, Tokyo 113-0033, Japan}

\date{\today}

  \maketitle

{\noindent\textbf {NOTE 1: Laue image of four-layer cuprates.}}
\vspace{0.5em}

\begin{figure}[htbp]
\includegraphics [clip,width=16.5cm]{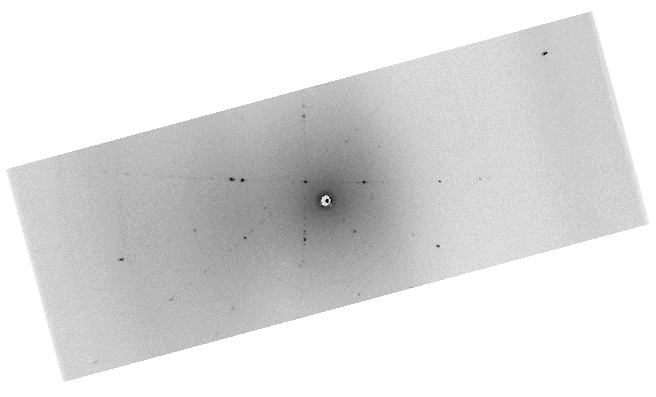}
\label{Laue}
\caption{\textbf{Laue image of a single crystal Ba$_2$Ca$_3$Cu$_4$O$_{8}$(F,O)$_2$.} 
Back-reflection Laue image of a four-layer cuprate sample, showing clear four-fold rotational symmetry without any sign of structural modulation.}
\end{figure}

Laue image of our single-crystal samples (Fig.~\ref{Laue}), taken on the \textit{ab}-plane in a back-reflection setup, shows a clear four-fold rotational symmetry. The absence of satellite reflections or spot splitting confirms that the crystal is free from structural modulation, which could otherwise complicate ARPES data in momentum space. The ring-shaped background intensities originate from the metal substrates used to glue the samples.\\

\textbf{NOTE 2: Error of Fermi pocket area (or carrier density $p$) determined by quantum oscillation data.}
\vspace{0.5em}

\begin{figure}[htbp]
\includegraphics [clip,width=16.5cm]{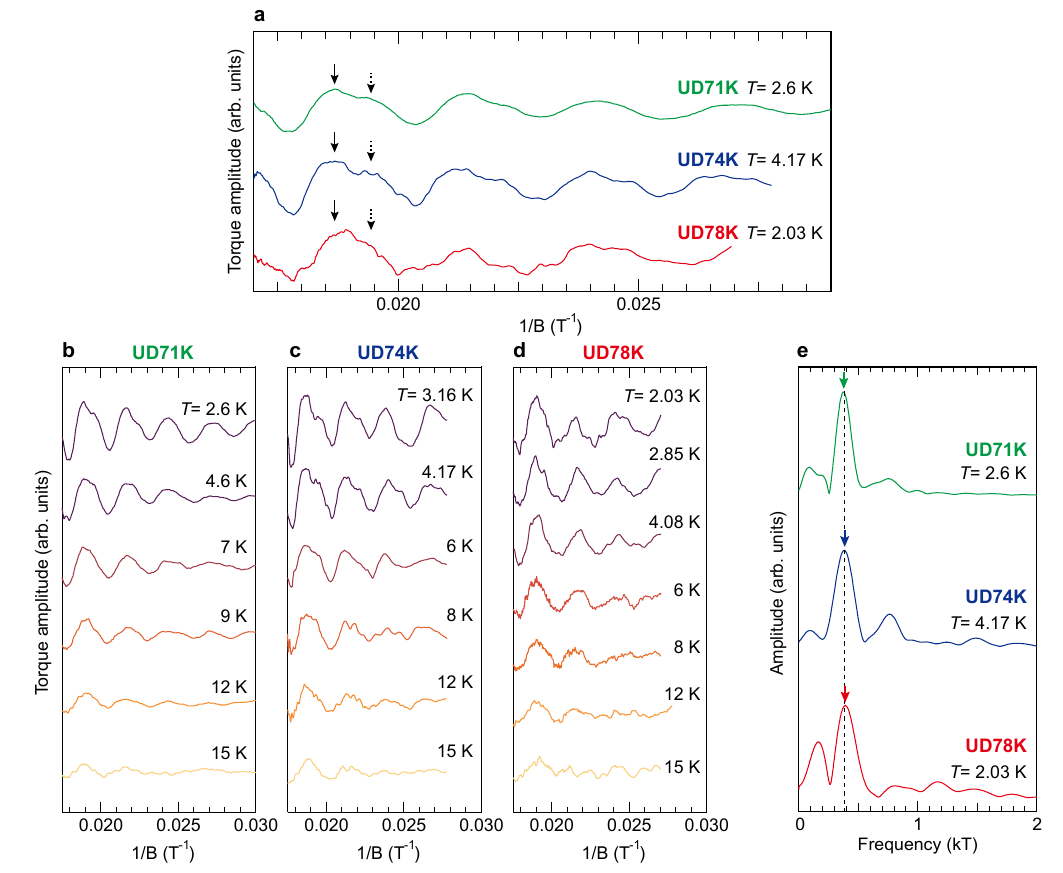}
\label{QuantumOscillation}
\caption{\textbf{Raw magnetic torque signals and their Fourier analysis.} 
\textbf{a,} Selected magnetic torque signals of UD71K, UD74K, and UD78K from (\textbf{b}-\textbf{d}). The presence of two oscillation components is indicated by black solid and dashed arrows.
\textbf{b}-\textbf{d,} Magnetic torque signals of UD71K, UD74K, and UD78K measured at various temperatures. A cubic background was subtracted from the raw data. The results after FFT analysis are shown in the main text Fig. 2\textbf{g}-\textbf{i}.
\textbf{e,} FFT spectra of the curves in (\textbf{a}), showing a single frequency peak near 380~T for all samples, confirming that the two components are nearly degenerate in frequency.}
\end{figure}

Figures~\hyperref[QuantumOscillation]{\ref*{QuantumOscillation}a}-\hyperref[QuantumOscillation]{\ref*{QuantumOscillation}d} present quantum-oscillation data of magnetic torque (de Haas-van Alphen effect) after background subtraction, which were used for fast Fourier transformation (FFT) analysis. The background signal was removed by fitting each raw torque signal with a cubic polynomial. The oscillation signals are plotted only for magnetic fields higher than the upper critical field (\emph{H}$_{c2}$), where superconductivity is fully suppressed.

As indicated by the solid and dotted arrows in Fig.~\hyperref[QuantumOscillation]{\ref*{QuantumOscillation}a}, the quantum oscillation data exhibit the main and shoulder-like secondary peaks. This behavior can be modeled as the sum of two sinusoid components, indicating that the quantum oscillation data include two distinct frequencies. These two frequencies likely originate from bilayer splitting between adjacent inner CuO$_2$ planes in the crystal structure (see Fig.~1a). However, the secondary frequency is nearly degenerate with the main one; the two are so close that the FFT yields only a single peak (Fig.~\hyperref[QuantumOscillation]{\ref*{QuantumOscillation}e}).

Note that the small peak at low frequency (150~T) in Fig.~\hyperref[QuantumOscillation]{\ref*{QuantumOscillation}e} are not intrinsic, since this minor peak arises as an artifact from the background subtraction. Other weak peaks at higher frequency (\emph{e.g.},~760~T) correspond to higher harmonics of the main oscillation frequency.

The error in estimating the Fermi pocket area (or carrier density $p$) was assessed from the stability of the main peak near 380~T in the FFT spectra obtained from quantum-oscillation data at various temperatures (Figs.~\hyperref[QuantumOscillation]{\ref*{QuantumOscillation}b}-\hyperref[QuantumOscillation]{\ref*{QuantumOscillation}d}). The full FFT spectra are shown in Figs.~1g-i of the main text. While the peak position shows slight temperature-dependent shifts, the variation is small within about 20~T, which roughly corresponds to the doping level $p$ of 0.3$\%$. We thus adopt $\pm 0.3\%$ of doping level as the experimental uncertainty. Note that such a small variation in the pocket area cannot be resolved in ARPES measurements due to the limited angular resolution (Fig.~\hyperref[QuantumOscillation]{\ref*{QuantumOscillation}f}).\\

{\noindent\textbf {NOTE 3: Estimation of the Fermi energy $\varepsilon_{\rm F}$ for UD71K, UD78K, and UD78K.}}
\vspace{0.5em}

In the main text, the Fermi energy ($\varepsilon_{\rm F} \sim 50$~meV) was estimated by fitting the band dispersion of UD71K.  
Here, we confirm that $\varepsilon_{\rm F}$ is comparable among the three samples: UD71K, UD74K, and UD78K, whose doping level differences are small within 0.6$\%$.

\begin{figure}[htbp]
\includegraphics [clip,width=16.5cm]{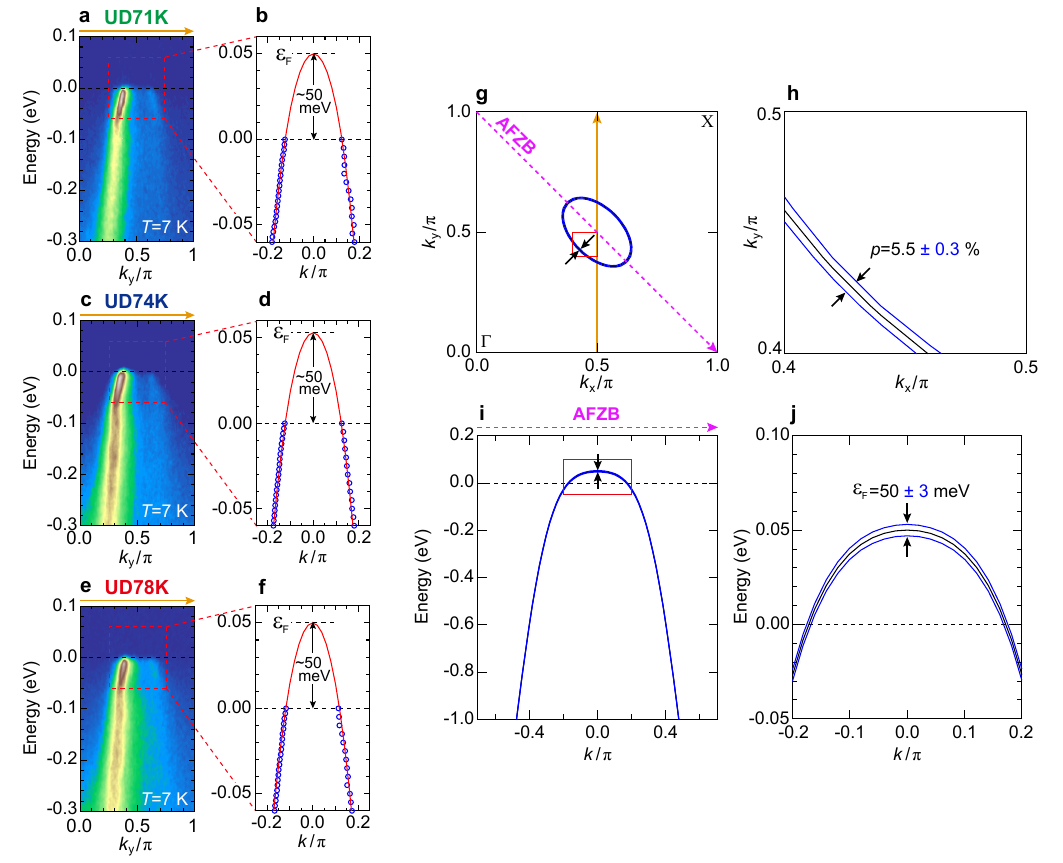}
\label{FermiEnergy}
\caption{\textbf{Estimation of the Fermi energy $\varepsilon_{\rm F}$ and its error.} 
\textbf{a,} ARPES band dispersion of the IP in UD71K, measured along the orange arrow in (\textbf{g}).
\textbf{b,} Band dispersion determined from the peak positions of the momentum distribution curves (MDCs) in (\textbf{a}). The Fermi energy $\varepsilon_{\rm F}$ (black arrow) is obtained by fitting the dispersion (red solid line) using a tight-binding model. 
(\textbf{c}-\textbf{f,}) Same measurement as (\textbf{a,b}), but for UD74K and UD78K, respectively.
(\textbf{g,}) Small Fermi pockets with three different carrier densities: $p=5.2\%$ (blue), $5.5\%$ (black), and  $5.8\%$ (blue). 
The tight-binding fit to the ARPES data with $p = 5.5\%$ was shifted in energy to simulate the Fermi pocket of $p=5.2\%$ and $5.8\%$.
(\textbf{h,}) Enlarged view of the red box region in (\textbf{g}).
(\textbf{i,}) Band dispersion for (\textbf{a}), plotted along the antiferromagnetic zone boundary (AFZB, pink dashed arrow in (\textbf{g})).  
(\textbf{j,}) Enlarged view of the boxed region in (\textbf{i}), showing that a doping variation of \(\pm 0.3\%\) corresponds to a shift in \(\varepsilon_{\rm F}\) of approximately \(\pm 3\)~meV (\(\varepsilon_{\rm F} = 50 \pm 3\)~meV).
}
\end{figure}

Figures~\hyperref[FermiEnergy]{\ref*{FermiEnergy}a}-\hyperref[FermiEnergy]{\ref*{FermiEnergy}f} show ARPES band dispersions near $E_{\rm F}$ for UD71K, UD74K, and UD78K.  
Fitting the occupied bands determines the top of the band, namely $\varepsilon_{\rm F}$, to be approximately 50~meV in all cases.  
The data for UD71K (Figs.~\hyperref[FermiEnergy]{\ref*{FermiEnergy}a},\hyperref[FermiEnergy]{\ref*{FermiEnergy}b}) are identical to those in the main text. This is consistent with the small doping difference ($\pm 0.3\%$) discussed in \textbf{NOTE 2}.
To estimate a possible variation of $\varepsilon_{\rm F}$ among the samples, we perform a tight-binding fit to the UD71K data and assume a rigid-band shift to simulate different doping levels.  

In Figs.~\hyperref[FermiEnergy]{\ref*{FermiEnergy}g},\hyperref[FermiEnergy]{\ref*{FermiEnergy}h}, we show Fermi pockets for a tight-binding band, shifted in energy to correspond $p = 5.2\%$ (blue), $5.5\%$ (black), and $5.8\%$ (blue).
The top of these bands is examined in Figs.~\hyperref[FermiEnergy]{\ref*{FermiEnergy}i},\hyperref[FermiEnergy]{\ref*{FermiEnergy}j}.  
$\varepsilon_{\rm F}$ varies only by $\pm 3$~meV.  
This variation is small enough to conclude that the abrupt increase of the pairing strength from UD71K to UD78K is not due to an error in the estimation of $\varepsilon_{\rm F}$, but an intrinsic feature in the four-layer cuprates.\\

{\noindent\textbf {NOTE 4: Tight-binding fitting to the Fermi pocket band.}}
\vspace{0.5em}

\begin{figure}[htbp]
\includegraphics [clip,width=8cm]{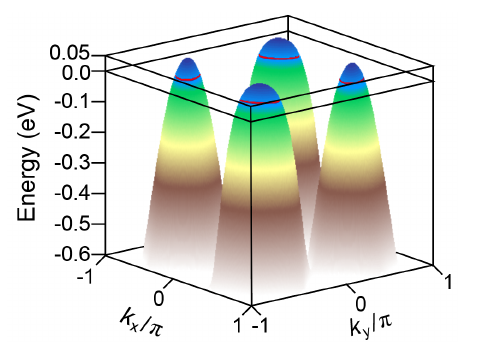}
\label{TB}
\caption{\textbf{Tight-binding band of the Fermi pocket.} 
Band dispersion determined by tight-binding fitting to our ARPES data.}
\end{figure}

In Figs. 1l and~\ref{FermiEnergy}, we perform a tight-binding fit to the ARPES data to extract the Fermi energy $\varepsilon_{\rm F}$. The same fitting parameters as those in the previous study on five-layer cuprates~\cite{Kunisada2020} are adopted here. Table~\ref{table1} summarizes the fitting functions and the corresponding coefficients used to reproduce the band dispersion in the current four-layer cuprate observed by ARPES. The tight-binding band is expressed as:
 \begin{equation}
     \epsilon_{\vec{k}} = \sum_{i=0}^{6} t_in_i
 \end{equation} The resulting band dispersion calculated using the parameters in Table~\ref{table1} is shown in Fig.~\ref{TB}.

 \begin{table}[hbtp]
    \centering
    \caption{Tight-binding functions and coefficients used in fitting to our ARPES data.}
    \renewcommand{\arraystretch}{1}
    \begin{tabular}{ccc}
    \hline\hline
    $\quad i\quad $ & $n_i$ & $\quad t_i \,\text{(eV)} \quad$\\ \hline
    0  & 1  & -0.545  \\
    1  & $\frac{1}{2}[\cos(k_x-k_y)+\cos(k_x+k_y)]$ & 0.1425  \\
    2  & $\cos(k_x-k_y)\cos(k_x+k_y)$ & -0.712  \\
    3  & $\frac{1}{2}\{ \cos[2(k_x-k_y)]+\cos[2(k_x+k_y)]\}$ & -0.0215 \\
    4  & $\frac{1}{2}\{\cos[2(k_x-k_y)]\cos(k_x+k_y)+\cos(k_x-k_y)\cos[2(k_x+k_y)]\}$ & 0.305 \\
    5  & $\cos[2(k_x-k_y)]\cos[2(k_x+k_y)]$ & -0.0955  \\
    6  & $\frac{1}{2}\{\cos[3(k_x-k_y)]+\cos[3(k_x+k_y)]\}$ & 0.045 \\ 
    \hline\hline
    \end{tabular}
    \label{table1}
    \end{table}
\newpage

{\noindent\textbf {NOTE 5: Estimation of doping levels in the outer planes.}}
\vspace{0.5em}

In the main text, the doping levels $p$ of IP, which form the Fermi pocket, were determined precisely by ARPES and quantum oscillation measurements. The variation of $p$ among the three samples (UD71K, UD74K, and UD78K) was found to be very small. Here, we determine the $p$ values of the outer planes (OP), which form the Fermi arc in these samples, and find that the doping level lies in the underdoped regime with similarly small variation among the samples, consistent with the result for the Fermi pockets in IP.

\begin{figure}[htbp]
\includegraphics [clip,width=16.5cm]{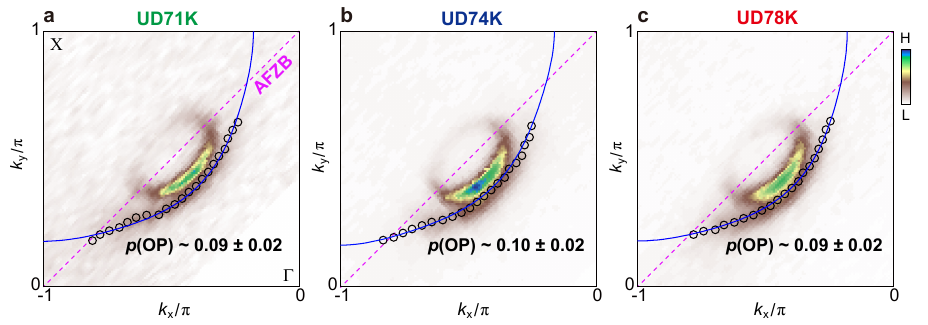}
\label{OPdoping}  
\caption{\textbf{Doping level estimation for the outer planes (OP) in three samples.} 
     \textbf{a,} Fermi surface mapping of UD71K measured by ARPES. The $k_{\rm F}$ points of the OP band (Fermi arc) are indicated by black circles. The blue line represents the tight-binding fit to the Fermi arc.
     \textbf{b,c,} Same measurements as (\textbf{a}), but for UD74K and UD78K, respectively.
     }
\end{figure}

Figures~\hyperref[OPdoping]{\ref*{OPdoping}a}-\hyperref[OPdoping]{\ref*{OPdoping}c} displays the Fermi surface mappings measured by ARPES for UD71K, UD74K, and UD78K, respectively. Aside from a small Fermi pocket in IP, a Fermi arc for OP is clearly visible. 

The doping level $p$ of OP is estimated by fitting the Fermi arc determined from ARPES spectra (open circles in Figs.~\hyperref[OPdoping]{\ref*{OPdoping}a}-\hyperref[OPdoping]{\ref*{OPdoping}c}) using a tight-binding model widely adopted for single- and double-layered cuprates \cite{Norman1995}. We find that the estimated $p$ values fall within the range of $\sim9\%$ to $\sim10\%$, indicating that all three samples are situated in the moderately underdoped regime. The small variation in $p$ across the samples is consistent with that observed in the IP. 
Notably, $9-10\%$ is also the doping range where the superconducting (SC) gap magnitude reaches its maximum in single- and double-layered cuprates. This further emphasizes the abnormality of the large gap observed in the Fermi pockets at a rather small doping level of $p\sim5.5\%$.\\

{\noindent\textbf {NOTE 6: Momentum dependence of the superconducting gap for all three samples.}}
\vspace{0.5em}

In the main text, we demonstrated the SC gap in detail only for UD78K, which has the largest SC gap among our samples. Here, we present the detailed SC gaps for the other samples (UD71K and UD74K), together with that of UD78K. 

\begin{figure}[htbp]
    \includegraphics [clip,width=16.5cm]{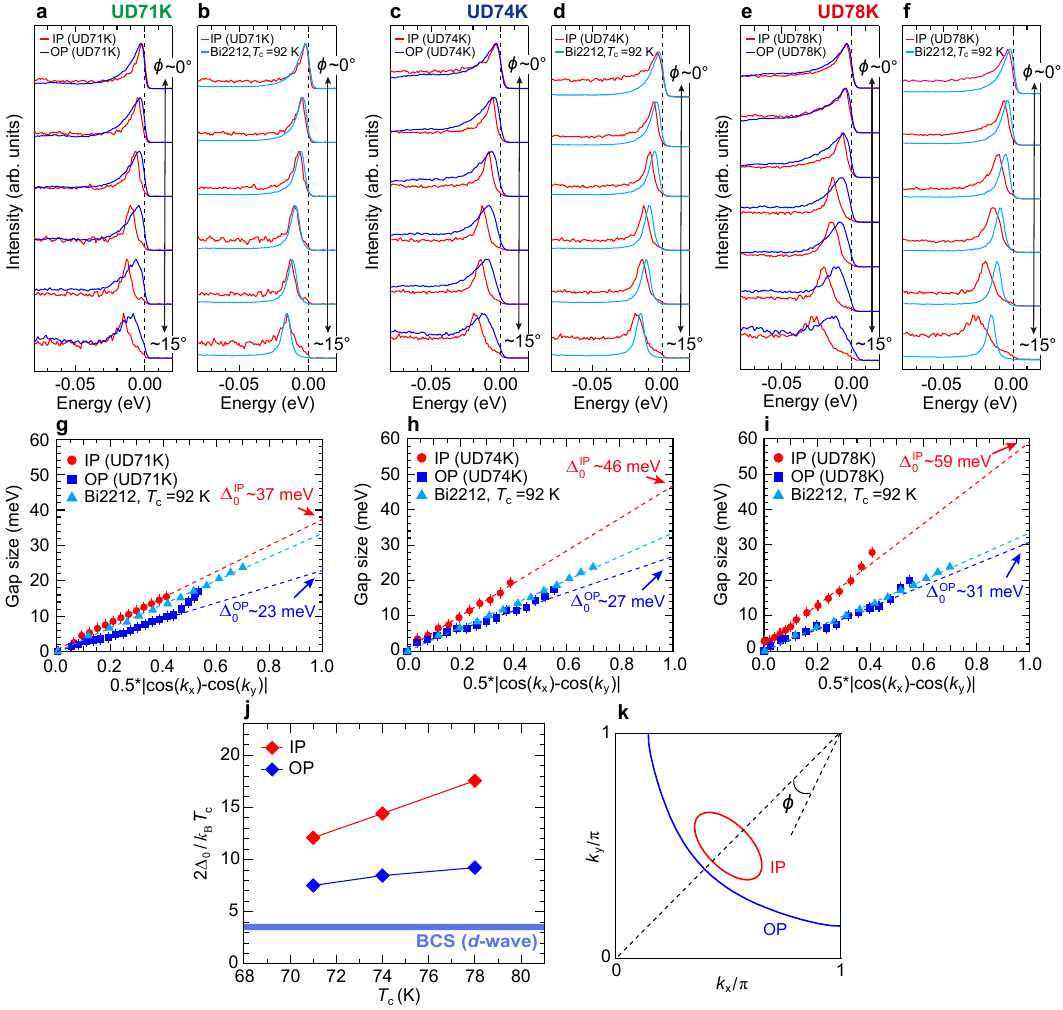}
    \label{gapSym}
    \caption{\textbf{Momentum-dependence of the superconducting gap for three samples.} 
    \textbf{a,} Comparison of the SC gap between the IP and OP in UD71K. EDCs taken at a similar Fermi surface angle $\phi$ are overlapped.
    \textbf{b,} Similar to \textbf{a}, but showing a comparison between the IP of UD71K and optimally doped Bi2212 \cite{Kondo_Bi2212}.
    \textbf{c}-\textbf{f,} Same measurements as in (\textbf{a,b}), but for UD74K and UD78K, respectively.
    \textbf{g,} Momentum dependence of the SC gap for each CuO$_2$ plane in UD71K and Bi2212, plotted as a function of $d$-wave form, $|\cos(k_x)-\cos(k_y)|/2$. The antinodal gap $\Delta_0$ is obtained by extrapolating the gap values near the nodal region. 
    \textbf{h,i,} Similar analysis to (\textbf{g}), but for UD74K and UD78K, respectively.
    \textbf{j,} Ratio $2\Delta_0$/\kbtc for each CuO$_{2}$ plane in our four-layer samples. 
    The thick horizontal blue line indicates the BCS weak-coupling prediction for the $d$-wave superconductivity~\cite{dwaveBCS}. 
    \textbf{k,} Schematic illustration of the Fermi surfaces for the IP and OP.
    }
\end{figure}

Figures~\hyperref[gapSym]{\ref*{gapSym}a}-\hyperref[gapSym]{\ref*{gapSym}f} show EDCs along the Fermi surface at the low temperature (10K), exhibiting the momentum dependence of 
the SC gaps in UD71K, UD74K, and UD78K. In each of the left panels (Figs.~\hyperref[gapSym]{\ref*{gapSym}a},\hyperref[gapSym]{\ref*{gapSym}c},and \hyperref[gapSym]{\ref*{gapSym}e}), the SC gaps of the inner plane (IP) and the outer plane (OP) are compared, whereas in each of the right panels, the SC gaps of inner plane (IP) and optimally doped Bi$_2$Sr$_2$CaCu$_2$O$_{8+\delta}$ (Bi2212) with the SC transition temperature ($T_{\rm c}$) of 92~K \cite{Kondo_Bi2212} are compared. It is observed that the IP gaps are much larger than the OP gaps in all three samples. In addition, the IP gap always exceeds the Bi2212 gap: while the degree of exceeding is small in UD71K, it becomes apparent in UD74K and significant in UD78K. 

These behaviors are demonstrated in Figs.~\hyperref[gapSym]{\ref*{gapSym}g}-\hyperref[gapSym]{\ref*{gapSym}i} more clearly by plotting 
the SC gaps as a function of $d$-wave form $|\cos(k_x)-\cos(k_y)|/2$. Here, the gap magnitudes were determined from the peak position of symmetrized EDCs (not presented). The extrapolated SC gap to antinode $\Delta_{0}$ is estimated by extrapolating the data around the nodal region as represented by the dotted line and arrow. Notably, the IP gap $\Delta_{0}^{\rm IP}$ shows a dramatic increase from 37 meV to approximately 60 meV, in stark contrast with the OP gap $\Delta_{0}^{\rm OP}$ showing rather moderate evolution. We emphasize that $\Delta_{0}$$\sim$60 meV is one of the largest SC gaps reported in cuprate research by ARPES. Notably, the OP is paramagnetic, whereas the IP exhibits an AF order. Furthermore, the doping level of the OP ($p=9\% \sim10\%$) is closer to the optimal doping than that of IP ($p=5.5\%$). Considering these seemingly unfavorable circumstances of IP, it is striking that the SC gap magnitude of IP is almost twice that of OP, reaching the largest gap in cuprates. This indicates that antiferromagnetic (AF) order does not compete with superconductivity, but rather coexists with it in an intimate relationship.

The extremely large gap magnitude is expected to yield a substantially large $\Delta_{0}$/\kbtc ratio in the small pocket. Figure~\hyperref[gapSym]{\ref*{gapSym}j} plots the $\Delta_{0}$/\kbtc ratio calculated for IPs and OPs in all three samples. Whereas the ratio for OP remains near 8.5, comparable to the previously reported value in underdoped Bi2212 \cite{Anzai2013},  the ratio for IP rapidly increases from $\sim$12 to $\sim$18, which far exceeds that of the BCS prediction~\cite{dwaveBCS}. This implies that the IP with the small pocket achieves the strong coupling, rapidly approaching the BEC regime.\\

{\noindent\textbf {NOTE 7: Much sharper and better-defined Bogoliubov quasiparticle peaks in the small pocket band than in the Fermi arc band}}
\vspace{0.5em}

\begin{figure}[htbp]
    \includegraphics [clip,width=16.5cm]{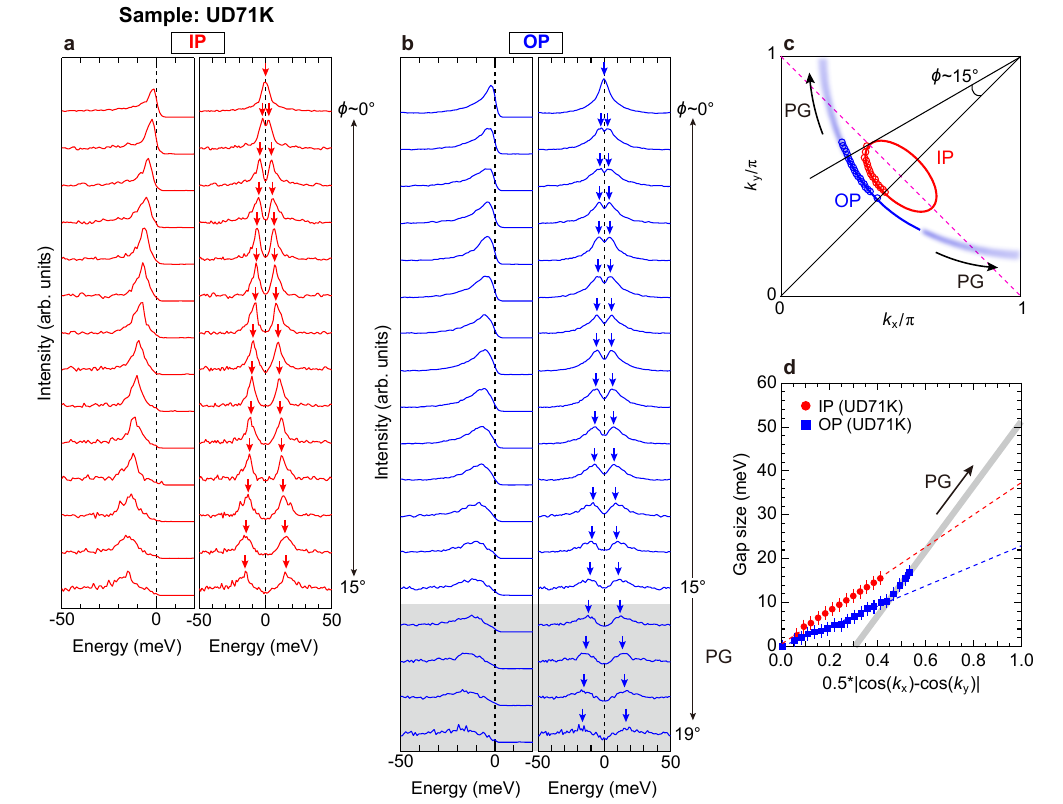}
    \label{UD71KEDC}    
    \caption{\textbf{Comparison of Bogoliubov quasiparticle peaks between the pocket band in IP and the arc band in OP.} 
    \textbf{a,} Raw (left panel) and symmetrized (right panel) EDCs of UD71K measured along the small Fermi pocket in the IP, from the nodal point to the pocket tip.
    \textbf{b,} Same as (\textbf{a}), but measured along the large Fermi surface in OP, from the nodal region toward the antinodal region. 
    The spectra in (\textbf{a,b}) were normalized to the integrated area over the energy range from -50 meV to 50 meV.
    They were then plotted with the same vertical offset.
    The spectra in a grey shaded area correspond to the spectra that deviate from the $d$-wave form, as shown in (\textbf{d}), indicating a pseudogap (PG) feature.
    \textbf{c,} Fermi surface of UD71K. Red and blue circles mark the $k_{\rm F}$ points corresponding to the spectra in (\textbf{a,b}), respectively. The blue-shaded line indicates the momentum region where broad spectra due to the pseudogap are observed. 
    \textbf{d,} Energy gap symmetry plotted with respect to the $d$-wave form. 
    The pseudogap with a larger energy scale than the SC gap causes a deviation from the simple $d$-wave function, as indicated by a gray-shaded line. 
 }
    \end{figure}

In \textbf{NOTE6}, we compared the SC gaps between the pocket band in IP and the arc band in OP. Here, we focus on the spectral line shape and compare it between IP and OP (Fig.~\ref{UD71KEDC}). 

Figures~\hyperref[UD71KEDC]{\ref*{UD71KEDC}a},\hyperref[UD71KEDC]{\ref*{UD71KEDC}b} show the energy distribution curves (EDCs) and those symmetrized about the Fermi level for IP and OP, respectively, for the UD71K samples. The Fermi momentum (\kf) points measured are indicated in Fig.~\hyperref[UD71KEDC]{\ref*{UD71KEDC}c} with circles. 
For a fair comparison, all spectra in Figs.~\hyperref[UD71KEDC]{\ref*{UD71KEDC}a},\hyperref[UD71KEDC]{\ref*{UD71KEDC}b} were normalized to their integrated area over the energy window of $[-50~\text{meV}, 50~\text{meV}]$. They are then plotted with the same value of a vertical offset. The spectra obtained from similar Fermi angles ($\phi$) are shown side by side to facilitate direct comparison between IP (Fig. \hyperref[UD71KEDC]{\ref*{UD71KEDC}a}) and OP (Fig. \hyperref[UD71KEDC]{\ref*{UD71KEDC}b}). It is evident that the Bogoliubov quasiparticle peaks (or the SC coherent peaks) in the Fermi pocket of the IP are significantly sharper, and thus better defined, than those in the Fermi arc of the OP. 

We further point out that, in the arc band of OP at $p\sim10\%$ (Fig.~\hyperref[UD71KEDC]{\ref*{UD71KEDC}b}), the pseudogap with a higher energy scale than the SC gap deviates the gap symmetry from a simple $d$-wave form.
The coherent peaks are almost absent at Fermi angles larger than 15$^\circ$, where only broad spectra are observed (see shaded regions in Figs.~\hyperref[UD71KEDC]{\ref*{UD71KEDC}b},\hyperref[UD71KEDC]{\ref*{UD71KEDC}c}). 
As a result, the Fermi surface length contributing to the superfluid density in the arc band of OP is comparable to that in the pocket band of IP. Combined with the finding that the SC gaps are much larger in the pocket than in the arc, this implies that the contribution of the pocket to the superfluid density, and hence to superconductivity, is likely greater than that of the arc. These findings suggest that the small pocket plays a dominant role in achieving the high-$T_{\rm c}$ in the four-layer cuprates. \\

{\noindent\textbf {NOTE 8: Bulk $T_{\rm c}$ predominantly determined by the IP, which forms the Fermi pocket band.}}
\vspace{0.5em}

We demonstrate that the SC transition temperature ($T_{\rm c}$) in our samples is predominantly determined by the IP, which hosts the Fermi pocket band. This is particularly important for the Uemura plot ($T_{\rm c}$/\emph{T}\textsubscript{F})~\cite{Uemura2004} of the four-layer cuprates. In the main text, we argue that $T_{\rm c}/\emph{T}\textsubscript{F}$ for IP reaches the theoretical upper limit of two-dimensional superconductivity ($T_{\rm c}/\emph{T}\textsubscript{F}$ = 0.125) \cite{Hazra2019}. This interpretation is based not only on the experimentally determined $\varepsilon_{\rm F}$ (=$k_{\rm B}T_{\rm F}$, where $k_{\rm B}$ is the Boltzmann constant and $T_{\rm F}$ is the Fermi temperature) for the pocket band in IP but also on the assumption that the $T_{\rm c}$ of the crystal is mainly decided by the superconductivity of IP, which forms the Fermi pocket. Here, we justify this assumption by comparing the evolution of the energy gap with temperature between IP and OP for UD78K, which has the largest SC gap among our samples.

\begin{figure}[htbp]
\includegraphics [clip,width=16cm]{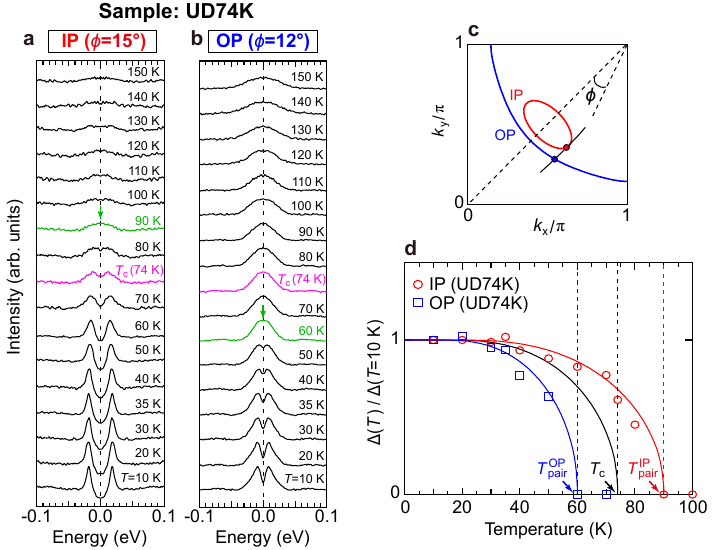}
\label{pairingT}
\caption{\textbf{Temperature dependence of the spectral gap in the pocket band (IP) and arc band (OP), and estimation of their gap-closing temperatures.} 
\textbf{a,} Symmetrized EDCs at the tip of the small Fermi pocket in IP ($\phi = 15^{\circ}$, red circle in (\textbf{c})), measured over a wide range of temperature. The spectrum at the $T_{\rm c}$ and gap-closing temperature is indicated in pink and green, respectively.
\textbf{b,} Similar to (\textbf{a}), but measured at the Fermi arc in OP ($\phi = 12^{\circ}$, blue circle in \textbf{c}).
\textbf{c,} Fermi surface of UD74K determined by the tight-binding fit. The black line indicates the measured momentum cut, crossing the $k_{\rm F}$ points where the EDCs in (\textbf{a,b}) were obtained.
\textbf{d,} The energy gap evolution with temperature for IP and OP. The gap sizes are normalized by those at the lowest measurement temperature ($T = 10$ K). The black curve represents a BCS-type gap function with $T_{\rm c}=$74 K. The red and blue curves represent BCS-type gap functions fitted to the data, which extrapolate the gap-closing temperatures ($T_{\rm pair}$) of each CuO$_2$ plane marked by coloured arrows.
 }
\end{figure}

Figures~\hyperref[pairingT]{\ref*{pairingT}a},\hyperref[pairingT]{\ref*{pairingT}b} show the temperature dependence of symmetrized EDCs measured at the same momentum cut (black line in Fig.~\hyperref[pairingT]{\ref*{pairingT}c}). Notably, the SC coherence peak in IP persists up to the bulk $T_{\rm c}$=74 K (pink curves), while a finite pairing gap remains above $T_{\rm c}$ and closes at $T = 90$ K ($T_{\rm pair}^{\rm ~IP}$ = 90 K; see the main text). On the other hand, the OP gap closes at the temperature lower than the bulk $T_{\rm c}$ (green curve in Fig.~\hyperref[pairingT]{\ref*{pairingT}b}). The spectral line shape may underestimate the gap-closing temperature; in such cases, an estimation of spectral weight filling at the Fermi level could be a more precise measure. Accordingly, we performed such an analysis, but the result remained unchanged. 

In Fig.~\hyperref[pairingT]{\ref*{pairingT}d}, we summarize the temperature dependence of the spectral energy gap for both CuO$_2$ planes (IP and OP). The data points are well fitted by a BCS-type gap function with the onset temperature $T_{\rm pair}$ estimated for each plane. $T_{\rm pair}^{\rm ~IP}$ is larger than the bulk $T_{\rm c}$, whereas $T_{\rm pair}^{\rm ~OP}$ is smaller than the bulk $T_{\rm c}$. We note that this corroborates the recent observations in other multilayer cuprates, suggesting that the clean IP determines the bulk $T_{\rm c}$ of the crystal \cite{Sun2025,Jeong2025}.

This set of results strongly supports our conclusion that the bulk $T_{\rm c}$ is predominantly determined by the IP. This conclusion is further supported by the following three observations discussed above:

\begin{enumerate}[label=(\arabic*), leftmargin=20pt, itemsep=2pt]
\item The superconducting gap magnitude ($\Delta_0$) in the IPs is much larger than that of OPs, even reaching approximately twice in the UD78K sample (Fig. \hyperref[gapSym]{\ref*{gapSym}} and \textbf{NOTE 6}).

\item The quasiparticle coherence peak is much sharper in IPs than that of OPs (Fig. \hyperref[UD71KEDC]{\ref*{UD71KEDC}} and \textbf{NOTE 7}). This implies that the pairs of IPs are more coherent with a longer lifetime.

\item The coherent portion of the Fermi surface that contributes to superconductivity is substantially reduced in the OPs because a large fraction of it is dominated by ill-defined pseudogap spectra. Consequently, the number of carriers participating in the superfluid density becomes comparable to that of a small pocket in the IPs (\textbf{NOTE 8}).
\end{enumerate}

Taken together, these findings suggest that, despite the lower overall carrier concentration in the IPs, their contribution to the total superfluid density is comparable to---or may even exceed---that of the OPs. This provides a justification for our classification of the four-layer cuprates on the Uemura plot in the main text.\\

{\noindent\textbf {NOTE 9: Comparison of pairing strength to other BCS-BEC crossover candidates.}}
\vspace{0.5em}

A key parameter describing the BCS-BEC crossover is the ratio of the pair size ($\xi \approx 1/\delta k_{\rm F}$) to 
 the interparticle spacing ($1/k_{\text{F}}$). This ratio can be expressed as $1/k_{\text{F}}\xi \approx \delta k_{\rm F}/k_{\rm F} \approx \Delta$/$\varepsilon_{\rm F}$. In the weak-coupling limit of the BCS superconductivity, a large number of pairs overlap with each other ($\xi \gg 1/k_{\text{F}}$). In contrast, the pair size shrinks (or the interparticle spacing gets large) and pairs become non-overlapping ($\xi \ll 1/k_{\text{F}}$) in the strong-coupling limit of BEC. Therefore, the ratio between these two characteristic lengths ($1/k_{\text{F}}\xi$ or $\Delta$/$\varepsilon_{\rm F}$) serves as a dimensionless parameter determining the proximity to the BEC limit, and the BCS-BEC crossover regime typically has a ratio of approximately unity.

The widely acknowledged pair size $\xi$ is the Ginzburg-Landau (GL) coherence length ($\xi_{\text{GL}} = \sqrt{\Phi_0 / 2\pi \mu_0 H_{c2}[1 - T/T_{c}]}$) and the Pippard coherence length ($\xi_{\text{Pippard}} = \hbar v_{\text{F}} / \pi \Delta$), where $\Phi_0$ is the flux quantum, $H_{c2}$ is the out-of-plane upper critical field, and $v_{\text{F}}$ is the Fermi velocity. The ratio $1/k_{\text{F}}\xi_{\text{Pippard}}$ is directly connected to the pairing strength $\Delta$/$\varepsilon_{\rm F}$ as follows, when a simple parabolic band structure is assumed:

\begin{equation}
\frac{1}{k_{\rm F}\xi_{\text{Pippard}}} = \frac{\pi}{2}\frac{\Delta}{\varepsilon_{\rm F}}
\end{equation}

In Table~\ref{Table2}, we present data collected from various superconductors reported to lie in the BCS-BEC crossover regime \cite{liu2012fese,zhang2019Fese,Lubashevsky2012,rinott2017,shruti2015,hashimoto2020,Nakagawa2021,suzuki2022,matbg_cao,matbg_oh} and compare those for four-layer cuprates determined by ARPES. The gap symmetry of cuprates is $d$-wave, in contrast to the $s$-wave symmetry in other superconductors on the list. Therefore, we used the SC gap and the Fermi wavenumber at the tip of the Fermi pocket as typical parameters for the four-layer cuprates. The coupling parameters ($\Delta$/\ef and $1/k_{\text{F}}\xi$) of the four-layer cuprates appear comparable to (or even larger than) those of other BCS-BEC crossover superconductors. These findings provide compelling evidence for the realization of a BCS-BEC crossover state in the clean IP of the four-layer cuprates.\\

{\noindent\textbf {NOTE 10: Determination of superconducting gap from the ARPES spectra.}}
\vspace{0.5em}

The superconducting gaps shown in this study were obtained by fitting the symmetrized EDCs taken at the $k_{\rm F}$ points to the phenomenological model as follows \cite{Norman1995}: 

\begin{equation}
    A(k_{\rm F}, \omega) = \frac{1}{\pi}\frac{\Sigma^{\prime\prime}}{(\omega-\Sigma^{\prime})^2+\Sigma^{\prime\prime2}} 
\end{equation}

where the self-energy ($\Sigma = \Sigma^{\prime}+i\Sigma^{\prime\prime}$) is expressed as:

\begin{equation}
     \Sigma(k_{\rm F}, \omega) = -i\Gamma_1+\frac{\Delta^2}{\omega+i\Gamma_0}
\end{equation}

Here, $\Gamma_1$ is the single-particle scattering rate and $\Gamma_0$ is the inverse-pair lifetime.

\begin{figure}[htbp]
\includegraphics [clip,width=16cm]{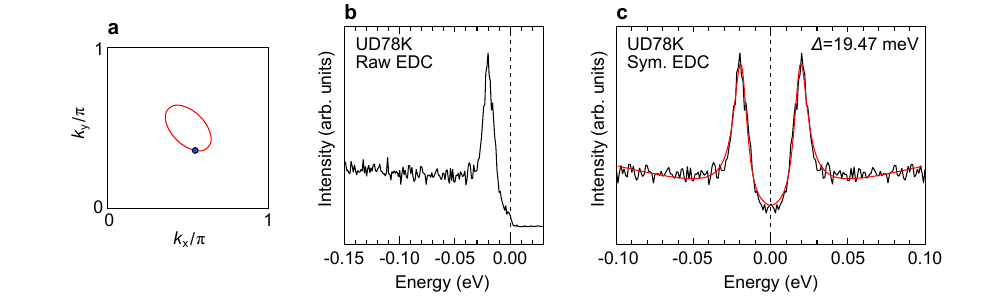}
\label{gapfitting}
\caption{\textbf{Superconducting gap fitting by phenomenological model.} 
\textbf{a,} Schematic illustration of IP in F0234. The blue circle indicates the $k_{\rm F}$ point where the data shown in (\textbf{b} and \textbf{c}) were acquired. 
\textbf{b,} \textbf{c,} Raw and symmetrized (Sym.) EDCs of UD78K in the superconducting state ($T=10$~K), taken at the momentum point shown in (\textbf{a}). The red solid curve indicates the fitting by the phenomenological model \cite{Norman1995}.
}
\end{figure}

In Fig. \hyperref[gapfitting]{\ref*{gapfitting}}, we present a representative gap analysis of our ARPES data. Figs. \hyperref[gapfitting]{\ref*{gapfitting}b} and \hyperref[gapfitting]{\ref*{gapfitting}c} are the raw and symmetrized (Sym.) EDCs of the UD78K sample, taken at the $k_{\rm F}$ point that is slightly away from the pocket tip (Fig. \hyperref[gapfitting]{\ref*{gapfitting}a}). After symmetrization, we fit the spectra with the phenomenological model, which matches the data well (Fig. \hyperref[gapfitting]{\ref*{gapfitting}c}). \\

{\noindent\textbf {NOTE 11: Temperature-dependence of Bogoliubov flat band.}}
\vspace{0.5em}

\begin{figure}[htbp]
\includegraphics [clip,width=16cm]{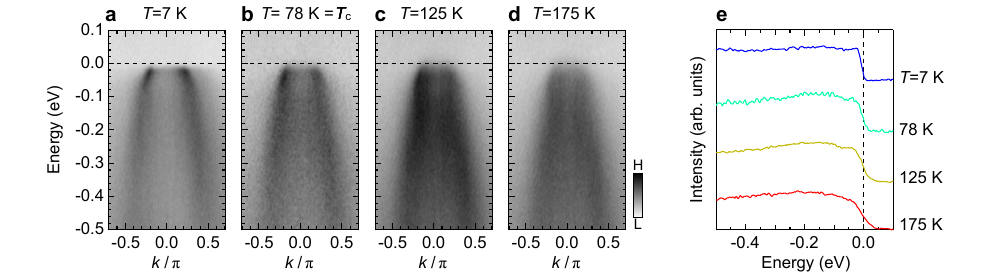}
\label{flatbandTdep}
\caption{\textbf{Temperature-dependence of the Bogoliubov flat band.} 
\textbf{a-d,} ARPES energy-momentum dispersions of UD78K along AFZB direction, measured at various temperatures. 
\textbf{e,} EDCs at the pocket center ($k=0$) from the dispersions in (\textbf{a-d}).
}
\end{figure}

In Fig. 4, we show the emergence of the Bogoliubov flat band in UD78K, a signature of BCS-BEC crossover. Here, we show the temperature-dependence of the Bogoliubov flat band. Figures \hyperref[flatbandTdep]{\ref*{flatbandTdep}a-d} present ARPES images of UD78K sample, taken along the antiferromagnetic zone boundary (AFZB) at several temperatures spanning below and well above (\tpair$\sim100$~K). The Bogoliubov flat band is clearly resolved at low temperature ($T=7$~K). With increasing temperature, the feature progressively broadens and becomes less distinct, and it is no longer discernible at $T=175$~K.

To further clarify this evolution, we plot temperature-dependent EDCs measured at the pocket center ($k=0$) in Fig.~\hyperref[flatbandTdep]{\ref*{flatbandTdep}e}. The low-temperature spectrum exhibits a pronounced peak-dip-hump structure, a characteristic signature of strong mode coupling commonly observed in cuprates. Although this feature persists into the normal state, it weakens at elevated temperatures and disappears around $T = 175$~K.

Notably, a similar flat band is not observed in the OPs, which are disordered due to their direct proximity to the dopant layer. This contrast supports that the Bogoliubov flat band observed in the UD78K sample is an intrinsic electronic property.\\

{\noindent\textbf {NOTE 12: Superconducting gap along AFZB.}}
\vspace{0.5em}

In Figs.~2e-f, we demonstrated the superconducting gap at the pocket tip, where it reaches its maximum value in the IPs of each sample. To further corroborate this gap magnitude, we examine the data along the AFZB cut, where $k_{\rm F}$ is unambiguously identified from the gap minimum.

\begin{figure}[htbp]
\includegraphics [clip,width=16cm]{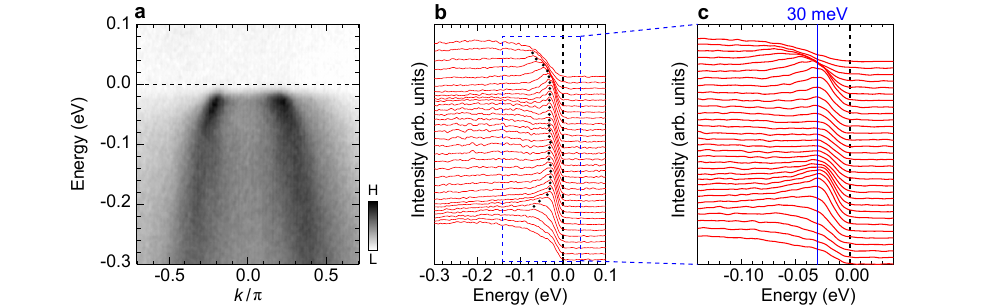}
\label{AFZBgap}
\caption{\textbf{Gap size of UD78K demonstrated by using momentum cut along AFZB.} 
\textbf{a,} ARPES dispersion of UD78K, measured along AFZB.
\textbf{b,} Set of EDCs, taken from $k/\pi = [-0.33, 0.36]$ in (\textbf{a}).
\textbf{c,} The same data as (\textbf{b}), but zoomed in near $E_{\rm F}$ (blue dashed box in (\textbf{b})). The gap magnitude is estimated to be $\sim30$~meV (blue line).
}
\end{figure}

Figures \hyperref[AFZBgap]{\ref*{AFZBgap}a} and \hyperref[AFZBgap]{\ref*{AFZBgap}b} show ARPES intensity map and the corresponding EDCs measured for UD78K below $T_{\rm c}$ ($T=7$~K):  the same data as in Figs. 4f, j.
To clarify the energy gap, we zoom into the spectra near $E_{\rm F}$ in Fig.~\hyperref[AFZBgap]{\ref*{AFZBgap}}.
The minimum gap, observed at $k_{\rm F}$, is estimated to be approximately $30$~meV (solid blue line in Fig.~\hyperref[AFZBgap]{\ref*{AFZBgap}c}) from the peak positions. These data confirm that a large superconducting gap of $\sim 30$~meV opens at the pocket tip.\\

{\noindent\textbf {NOTE 13: $d$-wave superconducting gap symmetry on the outer side of the small Fermi pocket.}}
\vspace{0.5em}

Previous ARPES studies on the five-layer system have established that the superconducting gap exhibits $d$-wave symmetry on both the inner and outer sides of the small Fermi pocket~\cite{Kunisada2020}. Here, we confirm that the same behavior holds for the present four-layer material.

\begin{figure}[htbp]
\includegraphics [clip,width=16cm]{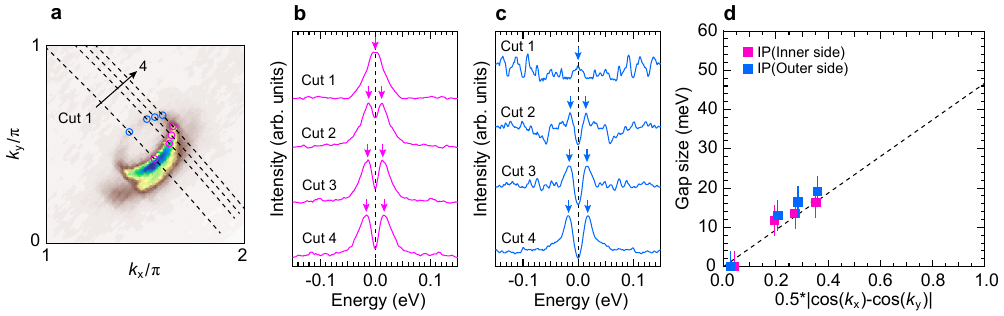}
\label{pocketgap}
\caption{\textbf{$d$-wave superconducting gap symmetry on the outer side of the small Fermi pocket.}
\textbf{a,} ARPES Fermi surface of UD74K. The dashed lines and colored circles represent momentum cuts and $k_{\rm F}$ points to acquire the data shown in (\textbf{b, c}).
\textbf{b,} Symmetrized EDCs of the inner side of the pocket in IP, from near-node (cut 1) to the pocket tip (cut 4), measured at the $k_{\rm F}$ points shown in (\textbf{a}) (magenta circles).
\textbf{c,} Similar to (\textbf{b}), but for the outer side of the pocket (blue circles in (\textbf{a})).
\textbf{d,} Momentum-dependence of the SC gaps shown in (\textbf{b, c}), plotted as a function of $d$-wave form. The dashed line is a guideline for the gap magnitude at the antinode ($\Delta_0$, see Fig. \hyperref[gapSym]{\ref*{gapSym}})
}
\end{figure}

Figures~\hyperref[pocketgap]{\ref*{pocketgap}b} and \hyperref[pocketgap]{\ref*{pocketgap}c} show representative EDCs of UD74K measured at the $k_{\rm F}$ points on the inner and outer sides of the pocket, respectively (marked by magenta and blue circles in Fig.~\hyperref[pocketgap]{\ref*{pocketgap}a}). Although the spectral intensity is much weaker on the outer side, reliable peak positions can still be extracted.

The extracted gap values follow a clear $d$-wave momentum dependence: the gap reaches its maximum near the pocket tip and decreases continuously toward the nodal direction, vanishing at the nodal cut. As shown in Fig.~\hyperref[pocketgap]{\ref*{pocketgap}d}, the gap magnitudes on the outer side quantitatively agree with those on the inner side within experimental uncertainty. These results confirm that the superconducting gap preserves $d$-wave symmetry over the entire pocket, including its outer region.\\

{\noindent\textbf {NOTE 14: Effective masses of the small Fermi pocket: comparison between ARPES and quantum oscillation measurements}}
\vspace{0.5em}

In the main text, we showed that the Fermi pocket size obtained by ARPES is fully consistent with that determined from quantum oscillation measurements. Here, we further compare the effective masses obtained from the two techniques.

In principle, ARPES is well-suited to evaluate the effective mass from the slope of the band dispersion. However, the mass variation observed in quantum oscillations is very small (see Figs.~1g-i), changing from $m^{\ast} \sim 0.82\,m_0$ in UD71K to $\sim 0.91\,m_0$ in UD78K (Figs.~1g-i). This corresponds to only a subtle modification of the band slope.

In cuprates, the intrinsic spectral linewidth is substantially broader than the change in dispersion slope associated with such a mass variation. As a result, resolving this small difference in effective mass directly from ARPES data is extremely challenging. To illustrate this limitation, we model a simple ellipsoidal band dispersion using the effective masses determined from quantum oscillation measurements and show that small mass variations are indistinguishable within the finite spectral linewidth. Nevertheless, we find that the overall dispersion agrees remarkably well with ARPES, demonstrating that ARPES and quantum oscillation measurements are quantitatively consistent. 

We consider a two-dimensional elliptical parabolic band $\varepsilon(\boldsymbol{k})$ as a function of momenta $k^{\rm Node}$ and $k^{\rm AFZB}$ setting axes along the nodal and AFZB directions:

\begin{equation}
    \varepsilon(\boldsymbol{k}) =\frac{\hbar^2}{2}\left(\frac{(k_{\rm Node})^2}{m_{\rm Node}}+\frac{(k_{\rm AFZB})^2}{m_{\rm AFZB}}\right), \quad \text{with}\quad \boldsymbol{k} = (k^{\rm Node}, k^{\rm AFZB}),
\end{equation}

where $m_{\rm Node}$ and $m_{\rm AFZB}$ are the effective masses along the nodal and AFZB axes. The constant-energy contour at $E=\varepsilon_{\rm F}$ is obtained from

\begin{equation}
    \frac{k^{\rm Node}}{k_{\rm F}^{\rm Node}} + \frac{k^{\rm AFZB}}{k_{\rm F}^{\rm AFZB}} =1,
\end{equation}

with

\begin{equation}
    k_{\rm F}^{\rm Node} = \sqrt{\frac{2m_{\rm Node}\varepsilon_{\rm F}}{\hbar^2}}, \quad k_{\rm F}^{\rm AFZB} = \sqrt{\frac{2m_{\rm AFZB}\varepsilon_{\rm F}}{\hbar^2}}.
\end{equation}

\begin{figure}[htbp]
\includegraphics [clip,width=16cm]{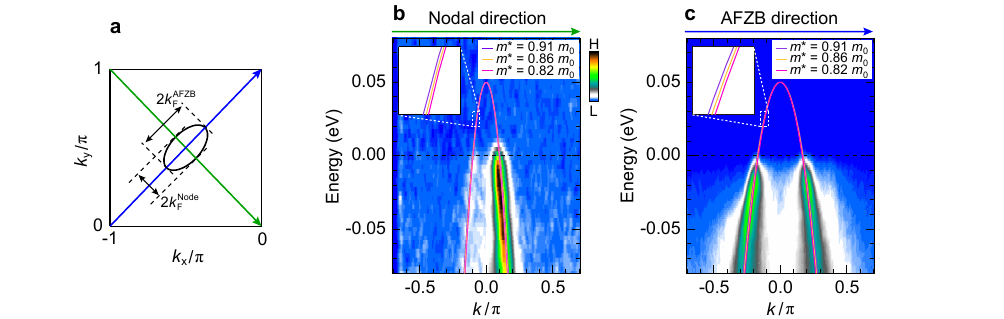}
\label{massplot}
\caption{\textbf{Effective masses of the small Fermi pocket: comparison between ARPES and quantum oscillation measurements.}
\textbf{a,} A Schematic illustration of the elliptical Fermi pocket. The green and blue arrows show the momentum cuts used in (\textbf{b, c}), respectively.
\textbf{b,} ARPES dispersion of UD74K, along the nodal direction (green arrow in (\textbf{a})). The colored curves are the parabolic band dispersions calculated by using the effective masses obtained from the quantum oscillation measurements. The inset figure (top left) is an enlarged plot of the parabolic bands (white dashed box), demonstrating that the expected slope change is very small.
\textbf{c,} Similar to (\textbf{b}), but for the AFZB direction cut (blue arrow in (\textbf{a})).
}
\end{figure}

Here, $k_{\rm F}^{\rm Node}$ and $k_{\rm F}^{\rm AFZB}$ correspond to the half size of the ellipsoidal Fermi pocket in the nodal and AFZB direction, respectively, as illustrated in Fig. \hyperref[massplot]{\ref*{massplot}a}.

The average mass along the Fermi surface $m^{*}$, or the cyclotron mass in quantum oscillation, is expressed as

\begin{equation}
    m^{*} = \sqrt{m_{\rm Node}m_{\rm AFZB}}
\end{equation}

Using  $k_{\rm F}^{\rm Node}$ and $k_{\rm F}^{\rm AFZB}$ obtained from the tight-binding fit to our ARPES data (Supplementary NOTE 4) and $\varepsilon_{\rm F}=50$~meV, we calculated the elliptical parabolic bands with three different masses: $m^{*}=0.82~m_0, 0.86~m_0$, and $0.91~m_0$ for UD71K, UD74K, and UD78K, respectively. For comparison, we overlay these curves on the ARPES dispersion for UD74K along the nodal and AFZB momentum cuts in Figs. \hyperref[massplot]{\ref*{massplot}b} and \hyperref[massplot]{\ref*{massplot}c}. 
Notably, the difference in the calculated band slopes for the three effective masses is very small (see the inset of Figs. \hyperref[massplot]{\ref*{massplot}b} and \hyperref[massplot]{\ref*{massplot}c}).  The expected variation is far smaller than the intrinsic spectral linewidth, so that it cannot be resolved by ARPES. Nevertheless, the calculated dispersions reproduce the ARPES data remarkably well, demonstrating that the ARPES and quantum oscillation results are quantitatively consistent.
\\

{\noindent\textbf {NOTE 15: Determination of errors in the pairing temperature.}}
\vspace{0.5em}

For UD71K and UD74K samples, the determination of \tpair is straightforward because the quasiparticle peaks clearly persist above $T_{\rm c}$. In these samples, we can determine the gap closing temperature (or \tpair) as the temperature where two peaks merge to one peak (Figs. 3a, b). Since the peak evolution is well resolved over multiple temperature points, the uncertainty in \tpair is small. Supplementary Fig. \hyperref[pairingT]{\ref*{pairingT}} presents the analysis for UD74K, where the gap-closing behavior is unambiguous and yields a relatively precise estimate of \tpair.

In contrast, UD78K does not show a clear peak-merging behavior at high temperatures. The spectra remain broadened due to a pseudogap that competes with superconductivity. In this case, distinguishing the pairing temperature \tpair from the pseudogap temperature $T^{*}$ requires a careful analysis of the temperature-dependent spectral loss at $E_{\rm F}$. Specifically, we identify \tpair as the temperature below which the spectral-weight loss upon cooling deviates from the pseudogap-derived linear behavior, following the procedure established in a previous ARPES study \cite{KondoNP}. 

\begin{figure}[htbp]
\includegraphics [clip,width=15cm]{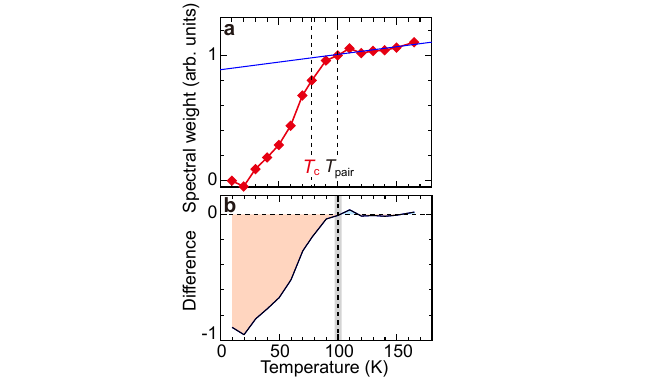}
\label{weight}
\caption{\textbf{Temperature-dependent spectral-weight-loss analysis for UD78K.}
\textbf{a,} Temperature-dependent spectral weight of UD78K, integrated near $E_{\rm F}$ (same plot shown in Fig. 3e). 
\textbf{b,} Difference plot of (\textbf{a}), obtained by subtracting the data points (red diamonds in (\textbf{a})) from the linear fit (blue line in \textbf{a}). The grey region corresponds to $\pm3$~K error range around \tpair.
}
\end{figure}

Based on this analysis, we estimate the uncertainty in \tpair for UD78K to be $\pm 3$~K. To clarify this point, we present the spectral-weight analysis for UD78K in Fig. \hyperref[weight]{\ref*{weight}}. Figure \hyperref[weight]{\ref*{weight}a} reproduces the data of Fig. 3e. In Fig. \hyperref[weight]{\ref*{weight}b}, we subtract the linear fit to experimental data in the high-temperature region from the overall data points. This procedure enhances the visibility of the deviation from the high-temperature linear behavior upon cooling, which we associate with the onset of pair formation~\cite{KondoNP}. The deviation occurs at approximately 100 K. Considering (i) the residuals of the linear fit and (ii) the finite temperature interval over which the deviation develops, we estimate an uncertainty of $\pm 3$~K, as indicated by the shaded region in Fig. \hyperref[weight]{\ref*{weight}b}.

\begin{sidewaystable}
    \centering
        \caption{Experimental data collected from various BCS-BEC crossover candidates. FeSe(ML)/STO, monolayer FeSe grown on the SrTiO$_3$ substrate; MATBG, magic-angle twisted bilayer graphene.}
            \renewcommand{\arraystretch}{1}
            \setlength{\tabcolsep}{4pt}
            \begin{tabular}{lccccccccc}
                \hline\hline
                Materials & \tc (K) & $\Delta$ & \ef (meV) & $\Delta$/\ef & \kf (nm$^{-1}$) & 1/\kf$\xi_{\text{Pippard}}$ & $\xi_{\text{Pippard}}$ (nm) & 1/\kf$\xi_{\text{GL}}$ & $\xi_{\text{GL}}$ (nm)\\
                \hline
               FeSe(ML)/STO \cite{liu2012fese,zhang2019Fese}& 38 & 15 & 60 & 0.25 & 2.3 & 1.2 & 0.36 & $\cdots$ & $\cdots$\\
                FeSe$_{x}$Te$_{1-x}$ \cite{Lubashevsky2012,rinott2017,shruti2015}& 12 & 3 & 6 & 0.5 & 1.2 & $\cdots$ & $\cdots$ & 2.21 & 0.38\\
                FeSe$_{1-x}$S$_{x}$ \cite{hashimoto2020}& 4 & 1.6 & 14 & 0.11 & $\cdots$ & $\cdots$ & $\cdots$ & $\cdots$ & $\cdots$ \\
                Li$_x$ZrNCl \cite{Nakagawa2021}& 15.9 & 4.08 & 11.3 & 0.36 & 0.52 & 3.42 & 0.56 & 5.41 & 0.36\\
                $\kappa$-(BEDT-TTF)$_4$HgBr \cite{suzuki2022}& 4.2 & $\cdots$ & $\cdots$ & $\cdots$ & 1 & $\cdots$ & $\cdots$ & 3 & 0.33\\
                4-layer cuprates (F0234)& 71 & 15 & 50 & 0.3 & 1.41 & 0.47 & 1.51 & $\cdots$ & $\cdots$ \\
                & 74 & 19 & 50 & 0.38 & 1.37 & 0.60 & 1.22 & $\cdots$ & $\cdots$ \\
                & 78 & 30 & 50 & 0.6 & 1.49 & 0.94 & 0.71 & $\cdots$ & $\cdots$ \\
                MATBG \cite{matbg_cao,matbg_oh}& 1 & 1.4 & 1.7 & 0.8 & $\cdots$ & $\cdots$ & $\cdots$ & $\cdots$ & $\cdots$ \\
                \hline\hline
            \end{tabular}
            \label{Table2}
\end{sidewaystable}
    
\clearpage

\clearpage

\renewcommand{\baselinestretch}{1.1}

